\documentclass[symmetry,article,submit,moreauthors,pdftex]{mdpi} 
\preto{\abstractkeywords}{\nolinenumbers}

\firstpage{1} 
\pubvolume{00}
\issuenum{1} 
\articlenumber{5} 
\pubyear{2020}
\copyrightyear{2020}
\Title{Pauli Crystals--Interplay of Symmetries}

\Author{Mariusz Gajda\orcidE{}, Jan Mostowski\orcidD{}, Maciej Pylak\orcidB{}, Tomasz Sowi\'nski $^*$\orcidA{} and~Magdalena~Za{\l}uska-Kotur\orcidC{}}

\AuthorNames{Gajda Mariusz, Mostowski Jan, Pylak Maciej, Sowinski Tomasz  and Zaluska-Kotur Magdalena}
\address[1]{%
Institute of Physics, Polish Academy of Sciences, Aleja Lotnikow 32/46, PL-02668 Warsaw, Poland}

\corres{Correspondence: Tomasz.Sowinski@ifpan.edu.pl}

\abstract{Recently observed Pauli crystals are structures formed by trapped ultracold atoms with the Fermi statistics. Interactions between these atoms are switched off, so their relative positions are determined by joined action of the trapping potential and the Pauli exclusion principle. Numerical~modeling is used in this paper to find the Pauli crystals in a two-dimensional isotropic harmonic trap, three-dimensional harmonic trap, and a two-dimensional square well trap. The~Pauli crystals do not have the symmetry of the trap--- the symmetry is broken by the measurement of positions and, in many cases, by the quantum state of atoms in the trap. Furthermore, the Pauli crystals are compared with the Coulomb crystals formed by electrically charged trapped particles. The~structure of the Pauli crystals differs from that of the Coulomb crystals, this provides evidence that the exclusion principle cannot be replaced by a two-body repulsive interaction but rather has to be considered to be a specifically quantum mechanism leading to many-particle correlations.}


\begin{document}
\section{Introduction}

Recent advances of experimental capabilities reached such precision that simultaneous detection of many ultracold atoms in a trap is possible~\cite{ReviewDetection,PhysRevA.97.063613}. This development paved the way to study positions of many trapped atoms subjected to the Fermi or the Bose statistics. What is even more important, simultaneous measurements of positions allow for the study of mutual correlations between particle positions. This is of particular interest for fermionic atoms, since the Pauli exclusion principle leads to nontrivial correlations between the position of atoms even if interactions between them are negligible. In this case, the Pauli crystals is the name coined for the spatial distribution of identical non-interacting fermions in a confined area. Such systems can be realized by e.g., optical traps. Particles with the Fermi statistics cannot be close to each other, because of the exclusion principle. Relative positions of particles, therefore, are determined by the joined action of the attractive trapping potential and the exclusion principle. These two mechanisms lead to nontrivial correlations of relative~positions.

In the context of the Pauli crystals, most attention was devoted to the study of a small number of fermions in a two-dimensional harmonic trap. The system was at zero temperature, hence in the quantum many-body ground state. For simplicity, it was assumed that the number of atoms matches the number of the quantum states in the lowest energy shells. It was determined by numerical modeling~\cite{2016GajdaEPL,2017RakshitSciRep,2017BatleAnnPhys,2019CiftjaAnnPhys,2019PyzhNJP,2020FremlingSciRep} and also in experiments with $^6$Li 
 atoms~\cite{2020HoltenARX} that indeed the atoms tend to take positions in the vicinity of vertices of non-trivial polygons, forming the so-called geometrical shells. Importantly, these shells do not have much in common with the energy shells. 

Relative positions of particles, hence the shape of the Pauli crystal, are directly related to inter-particle correlations present in a particular many-body quantum state. The standard approach to many-body correlations with the help of the high order correlation functions has, however, serious~shortcomings. The correlation functions depend on many arguments, as many as the number of particles, so their presentation and analysis is complicated. On the contrary, as argued in~\cite{2016GajdaEPL}, the shape of the Pauli crystal gives a natural way of the study and visualization of the many-body correlations between particle positions. 

The Pauli crystals exhibit unexpected symmetries. Some of them are simple and follow from the symmetry of the trapping potential. Some others can be linked to the closest packing principle. In~most cases, however, none of the crystal symmetry can be explained in such way. Typically, the~Pauli crystals show structures that do not have simple relations with the close packing nor with the shape of the potential, but rather follow from an interplay of many symmetries that leads to new geometric~structures.

The Pauli crystals are not the only structures formed by trapped particles. A lot of attention has been paid to the Coulomb crystals formed by classical charged particles confined in a trap~\cite{1985MostowskiAPPA, 1987Walther, 1987Wineland} or the Wigner crystals formed by charged particles in a uniform oppositely charged background~\cite{1934WignerPhysRev,2009Wigner,2012MonarkhaReview,2020ReganNature, 2019Wigner1d}. In both cases the equilibrium is determined by the mutual action of the attractive trapping forced by the potential or background and the repulsive Coulomb interaction between the particles; these structures were observed in many laboratories (see e.g.,~\cite{2012MonarkhaReview,2015Thompson} and references therein). When comparing these classical structures to the Pauli crystals, it should be noted that the electrostatic interaction between singly charged ions leads to much larger distances between the particles than the exclusion principle in the case of the Pauli crystals. Thus, the requirements for the observation of the Pauli crystal, temperature and spatial resolution, are much more strict than in the case of Coulomb crystals. Classical~mechanics is sufficient for a theoretical description of ions forming a Coulomb crystal in a trap since they are separated by a high potential barrier and their wave functions do not overlap. This~should be opposed to the Pauli crystal, where the overlap of wave functions of individual particles is the leading factor that determines the structure. 

In the present paper, we discuss the geometric structures of the Pauli crystals in case of a two-dimensional isotropic harmonic trap and also in a spherically symmetric harmonic trap in three spatial dimensions. Similarities and differences between the Pauli crystals and Coulomb crystals are discussed. In addition, we discuss cases when the particles do not form closed shells, hence~the symmetry of the trap is broken by the quantum state. Finally, we discuss the case of particles in a two-dimensional square trap. Here the natural symmetry of space is broken by the potential. The~symmetry of the trap influences in a significant way the shape of Pauli crystals and leads to an interesting interplay of the symmetry of the square and the natural symmetry of fermions.

\section{Pauli Crystals}
In this section, we will recapitulate basic properties of the Pauli crystals, the way they are simulated numerically, and measured in real experiments. For this purpose, let us consider a relatively small number $N$ of fermions, i.e., half-integer-spin particles with all their spins permanently polarized. For simplicity we will consider here a two dimensional harmonic isotropic trap. The Hamiltonian of the system in the second quantization formalism reads
\begin{equation}
  {\cal H} = \int\!\!\int\!\mathrm{d}x\,\mathrm{d}y\,\hat{\Psi}^\dagger(x,y)\left[-\frac{\hbar^2} {2m}\nabla^2+\frac{m\Omega^2}{2}(x^2+y^2)\right]\hat{\Psi}(x,y).
\end{equation}
where $m$ and $\Omega$ denote the mass of the particles and the oscillation frequency of these particles in the trap, respectively. Fermionic field operators $\hat\Psi(x,y)$ fulfill the standard fermionic anti-commutation~relations
\begin{subequations}
\begin{align}
  \{\hat\Psi(x,y),\hat\Psi^\dagger(x',y')\}&=\delta(x-x')\delta(y-y'), \\
  \{\hat\Psi(x,y),\hat\Psi(x',y')\}&=0
\end{align}
\end{subequations}
and can be decomposed in the basis formed by eigenstates of the single-particle Hamiltonian as
\begin{equation}
  \hat{\Psi}(x,y) = \sum_{n} \sum_{m} \hat{a}_{nm} \varphi_{nm}(x,y).
\end{equation}

Here, the wavefuntion $\varphi_{nm}(x,y)$ denotes the energy eigenstate with excitation $n$ and $m$ in the $x$ and $y$ spatial direction, respectively. The corresponding single-particle energy is equal to \mbox{$E_{nm}=\hbar\Omega(n+m+1)$}, thus states with the same sum $n+m$ belong to the same single-particle energy shell. The state $\varphi_{00}(x,y)$ denotes the single-particle ground-state. 

In principle, simultaneous measurement of $N$-particle positions is possible. According to the rules of the quantum mechanics the results are random, i.e., repeated measurements give different locations of the particles after each measurement. The measured positions are, however, strongly correlated and the probability of finding a given set of positions is determined by the state dependent probability~density
\begin{equation}
  \rho^{(N)}(\boldsymbol{r}_1,\ldots,\boldsymbol{r}_N) = \langle\mathtt{S}|\hat{n}(\boldsymbol{r}_1)\cdots\hat{n}(\boldsymbol{r}_N)|\mathtt{S}\rangle,
  \label{distribution}
\end{equation}
where $\hat{n}(\boldsymbol{r})=\hat{\Psi}^\dagger(\boldsymbol{r})\hat{\Psi}(\boldsymbol{r})$ is the single-particle density operator, and $|\mathtt{S}\rangle$ denotes the quantum state of the system. In this work we focus only on pure many-body states of the system. Generalization to mixed states, described by density matrix operators, is straightforward~\cite{2017RakshitSciRep}.

A natural way to discuss correlations between the particle positions is to use the hierarchy of correlation functions. The lowest of these functions give the one-particle density function $\rho^{(1)}(\boldsymbol{r})=\langle\mathtt{S}|\hat{n}(\boldsymbol{r})|\mathtt{S}\rangle$--- the probability density of finding a particle at a given position. The second correlation function $\rho^{(2)}(\boldsymbol{r}_1,\boldsymbol{r}_2)=\langle\mathtt{S}|\hat{n}(\boldsymbol{r}_1)\hat{n}(\boldsymbol{r}_2)|\mathtt{S}\rangle$ determines joined probability density of finding two particles at given positions, {\it etc.} Higher correlation functions, as functions of many arguments, are~difficult to visualize and to discuss. Particularly, this is the case when the highest correlation function $\rho^{(N)}(\boldsymbol{r}_1,\ldots,\boldsymbol{r}_N)$ is considered. 

The study of the correlation functions is not the only way to discuss the particle positions and correlations between them.   
An alternative procedure was suggested recently~\cite{2016GajdaEPL,2017RakshitSciRep}, it allows capturing and visualizing geometrical correlations between particle positions encoded in the total probability density $\rho^{(N)}(\boldsymbol{r}_1,\ldots,\boldsymbol{r}_N)$. The procedure bypasses difficulties caused by the correlation functions. It~begins with the study of the many-particle distribution function and finds the location of particles with the largest probability. These positions form a unique pattern. Next, one performs a sequence of measurements of particle positions in repeated experiments. In the numerical simulation, these~measurements are replaced by a random choice of positions, with the probability distribution given by the modulus square of the wave function. After each such measurement, the particle positions are shifted so that the measured center of mass is moved to the center of the trap, and rotated by such an angle that the particles come as close as possible to the pattern. These are symmetry transformations that do not change the measured relative positions of particles. After many measurements and transformations described above one obtains a specific configuration density function ${\cal C}(\boldsymbol{r})$ formed by all registered positions. The resulting geometrical structure is called the Pauli crystal. Details of this procedure in the theoretical framework can be found in~\cite{2016GajdaEPL,2017RakshitSciRep} and in the experimental realization in~\cite{2020FremlingSciRep}.

It should be noted that the geometric structures shown explicitly here are also hidden in the standard correlation functions. The applied procedure of handling the data obtained from many measurements of particle positions allows for a simple visualization of the resulting shapes. 

\section{Few  Particles  in  One Dimension}
Let  us  first  look  at  the  one-dimensional case  of  non-interacting  particles in a harmonic trap.  In  this case the  wave function  of the ground state of  the  system $|\mathtt{S}\rangle$ is a simple Slater determinant of $N$ lowest one-particle orbitals $\varphi_{n}(x)$. Each  orbital  is  given by the standard harmonic-oscillator wave function:
\begin{equation}
\varphi_n(x)={\cal N}_n {\cal H}_n (x) e^{-x^2/2}
\end{equation}
where ${\cal N}_n=(2^n n! \sqrt{\pi})^{-1/2} $ is the  norm  and  ${\cal H}_n(x)$ is  $n$-th Hermite  polynomial.  The  position  $x$ is  expressed  in  the normal harmonic  oscillator  units $a= \sqrt{\hbar/m\Omega} $,  where $m$ is  the  particle mass,  and  $ \Omega $  is  the  frequency  of  the trap. Please note  that  there  is  no other length scale in  this system, this length scale determines the  system geometry. In  a non-interacting many-particle state  each fermion occupies  one  state,  and therefore we  have   $N$-particle probability   density  function in  the  form: 
\begin{equation}
\label{prob_1D}
\rho^{(N)}(x_1,...,x_N)=({\cal N}_0...{\cal N}_{N-1})^2 \prod_{i < j} (x_i-x_j)^2  e^{-\sum_{i=1}^N x_i^2}
\end{equation}   

The  function above describes the probability density that  positions of $N$ particles  measured   simultaneously  in  one  measurement   will  be given  by ${x_1, ...x_N}$. Please note  that  the  same  function  describes  Bose particles in  1D with  infinitely large, delta-like  repulsion,  the Girardeau model~\cite{1960Girardeau}. It  means  that  the results for  such system  will  be  the  same as these  for  Fermion  system.
 
\begin{figure}[H]
  \centering
  \includegraphics[height=4.5cm]{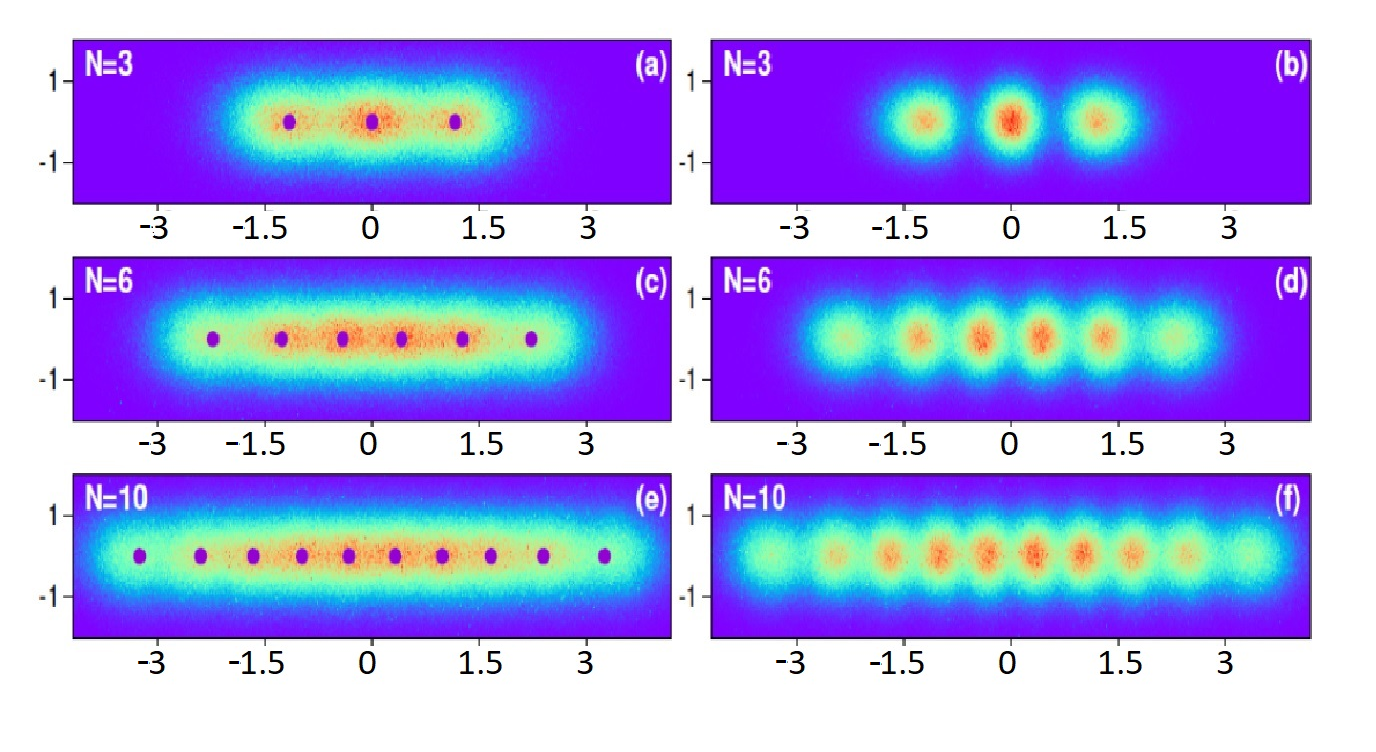}
\caption{The one-particle density function $\rho^{(1)}(\boldsymbol{r})$ (left) and the corresponding Pauli crystal (the~ configuration density function ${\cal C}(\boldsymbol{r})$) (right) for $N=3,6$ and  10 fermions confined in one  dimensional harmonic trap. Black dots correspond to the most probable configuration of particles (the~pattern).  All~positions are scaled with respect to the natural oscillator length unit $\sqrt{\hbar/m\Omega}$.} 
  \label{Fig0}
\end{figure}

To  analyze the  probability  distribution (\ref{prob_1D})  first  we  will find the particle pattern.  Configurations  that  correspond  to  the maximal  value   of the probability  density  function  (\ref{prob_1D}) for  small  number  of  particles are  relatively easy to  find. For  two  particles  the maximum   of function  (\ref{prob_1D}) is  for  \mbox{$\{x_1, x_2\}= \{-1/\sqrt{2},1/\sqrt{2}\}$}, three particles will most  probably be  found at  $\{x_1, x_2,x_3\}= \{-\sqrt{3/2},0,\sqrt{3/2}\}$ and four  particles at  points   $\{x_1, x_2,x_3,x_4\}= \{-1.64,-0.523, 0.523, 1.64\}$. All values  are  given  in the normal  harmonic  oscillator  unit $a$. In  such  a way  we  have  found  the  unique pattern  for  each  of   these systems.  It~ consist particle positions for  which the function (\ref{prob_1D}) has the maximum.  This  pattern  will  be the basis,  all~measured sets of positions will be compared to this basic set. The  question  is  if  such  procedure leads  to  distinguishable spots in the general case. In  other  words  if, according  to  the probability  distribution~(\ref{prob_1D}),  the  dispersion  around  the well-defined  maximum  is small of  large.  We  estimated this  quantity  for  each of  the examples listed above by the dispersion  of  position  of  one  particle,  assuming  that all other particles occupy positions given  by  the  
pattern  of  maximal probability. We  calculated  this dispersion using  conditional  probability $\rho_{cond}=\rho(x|x_2, ..., x_N)$. It  turned out  that  in  each  calculated  case  the dispersion of  the  particle position $\sigma=\sqrt{\langle(x-x_1)^2\rangle}$ was  the  same,  equal  to  $1/\sqrt{2}$ in the oscillator  units. Interestingly,  this  value was  the  same  for  each  position  for  each  system,  so it seems  to  be  some  universal  value. The  most  important  result is  that  this  value  is equal to one half  of  the  distance  between the  particles in  two  particle system  and  a little more  than one half  for  systems  with more  particles. It~means that the spots around  the  main  particle position should  be well  separated  from each other. This is illustrated in  Figure \ref{Fig0},  where  the single particle density  at  the left  side  is  compared  with  the   results  of  the numerical analysis of  many-particle density  function,  for  systems  of  a few particles  in  a quasi-one dimensional trap (harmonic trap elongated along one  direction with  aspect  ratio  1 to 10). It~can  be  seen  that  the  distances  between particles decrease with  number  of  particles,  but  the  individual  spots  around  each  particle position  can be clearly seen.\unskip
\unskip\vspace{-6pt}

\section{Closed Shells }
  The many-body wave function of the state in the two dimensional trap $|\mathtt{S}\rangle$ is a simple Slater determinant of $N$ lowest one-particle orbitals $\varphi_{nm}(\boldsymbol{r})$,  dependent  on  two  quantum  numbers $n$ and~$m$. In  this  case, due to the  degeneracy of orbitals with the same total number of excitations $n+m$, one~distinguishes two different scenario: {\it (i)} the number of particles can be such that all one-particle states of the same energy are occupied (closed energy shell scenario) and the ground state is non-degenerate, or {\it (ii)} the number of one-particle states of given energy does not fit the number of particles and then the many-body ground state is not unique (energy shells are not closed).

\begin{figure}[H]
  \centering
  \includegraphics[height=6cm]{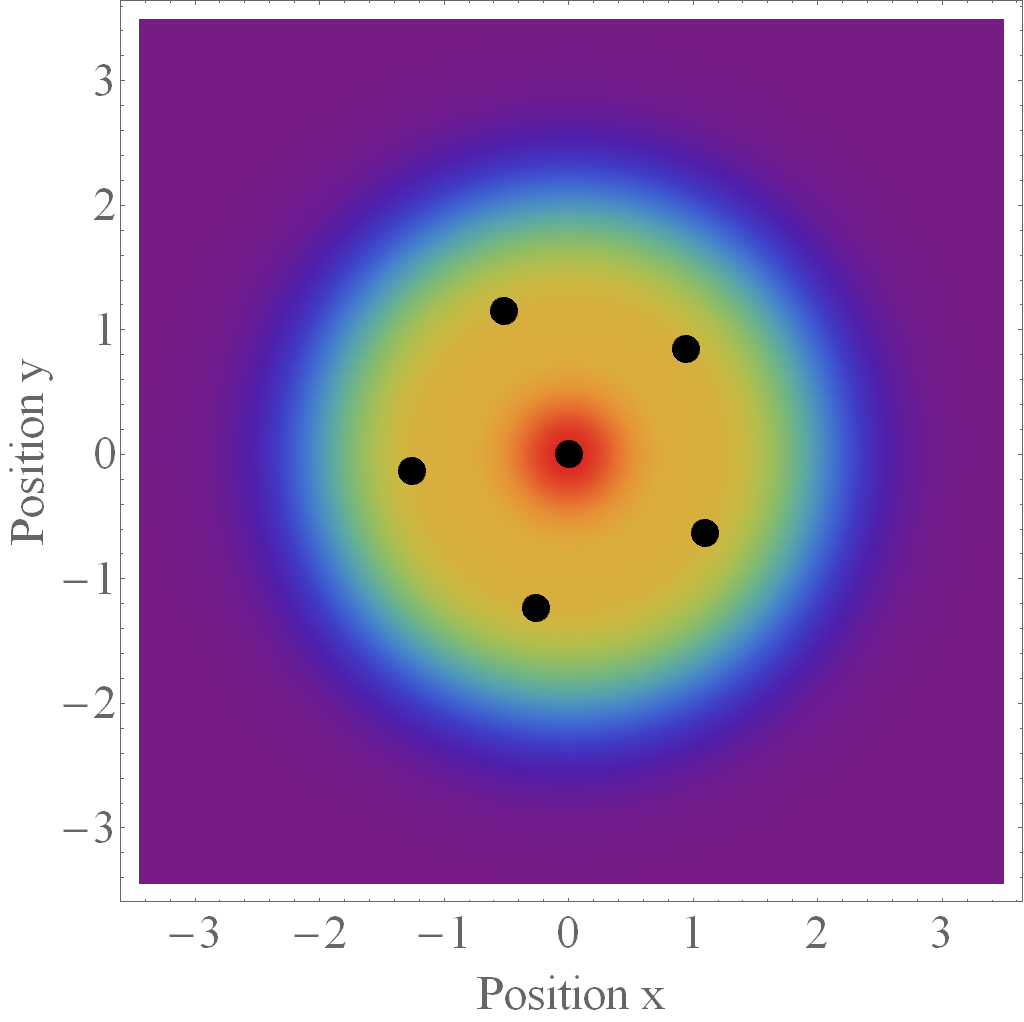}
  \includegraphics[height=6cm]{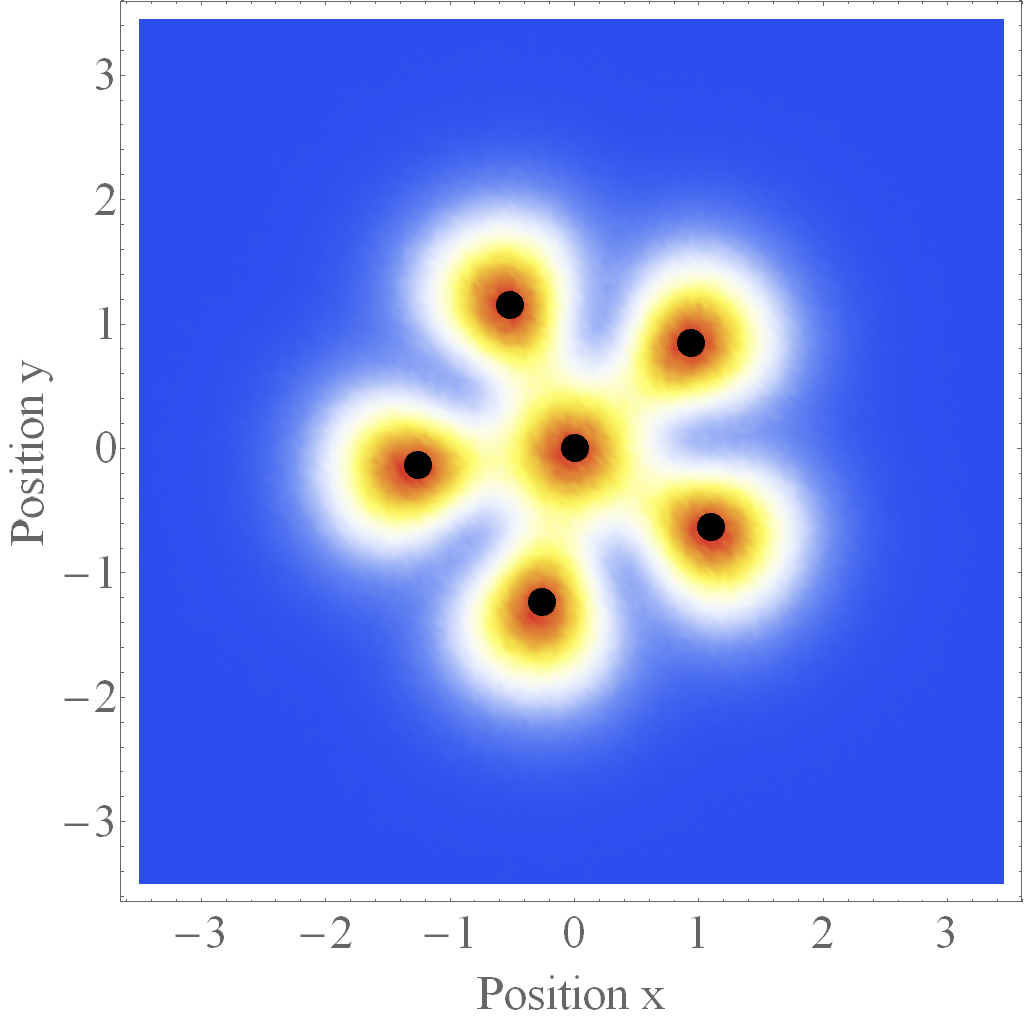}
\caption{The one-particle density function $\rho^{(1)}(\boldsymbol{r})$ (left) and the corresponding Pauli crystal (the~configuration density function ${\cal C}(\boldsymbol{r})$) (right) for $N=6$ fermions confined in an isotropic harmonic trap. Black dots correspond to the most probable configuration of particles (the pattern). Two~geometrical shells containing one and five fermions are clearly visible. The color in the 2D plot gives the configuration density function ${\cal C}(\boldsymbol{r})$. Please note that the structure of energy shells is different. All~positions are scaled with respect to the natural oscillator length unit $\sqrt{\hbar/m\Omega}$.} 
  \label{Fig1}
\end{figure}

In this section, we will discuss the closed energy shells scenario. In this case, the wave function is uniquely determined by the total energy and in two dimensions and is realized whenever \mbox{$N=1, 3, 6, 10,$~{etc.}} particles are considered. These numbers originate from the $n$-fold degeneracy ($n=1, 2,\cdots$) of each energy state. As an instructive example, let us examine the case of $N=6$ particles which occupy the three lowest energy shells. After performing extraction of the Pauli crystal and obtaining the configuration density function ${\cal C}(\boldsymbol{r})$ one finds that the number of geometrical shells differs from the number of energy shells (see Figure~\ref{Fig1} reproduced from~\cite{2016GajdaEPL}). As opposed to the energetic structure of the state, only two geometric shells are clearly seen. The inner shell consists of one particle, while five particles form the outer shell. It is very important to note that, although the quantum state describing particles in the case of closed shells is rotationally invariant, the Pauli crystal violates the rotational symmetry of the trap. The rotational symmetry is broken by the measurement and an arbitrary choice of the pattern orientation during the data analysis. Equally well the whole shape could be rotated by an arbitrary angle.
A naive intuition may suggest that the particles should have a tendency to be located at vertices of a regular hexagon. This is not the case, the probability density function does not distinguish such configuration. One can also look at the fivefold symmetry of the outer shell from another perspective and consider it as something natural. For example, it is well-known that the Coulomb crystals (electrically charged particles in an external harmonic trap) exhibit the same shape. In fact, the Coulomb interactions are not specific for the fivefold symmetry and any repulsive two-body interactions depending on the relative distances only lead to the same shape of the particle positions. This is because such systems realize the principle of the closest packing regardless of the details of interactions. The Pauli crystal formed by six particles exhibits the same behavior even though there are no two-body interactions between the particles.

\begin{figure}[H]
  \centering
  \includegraphics[height=6cm]{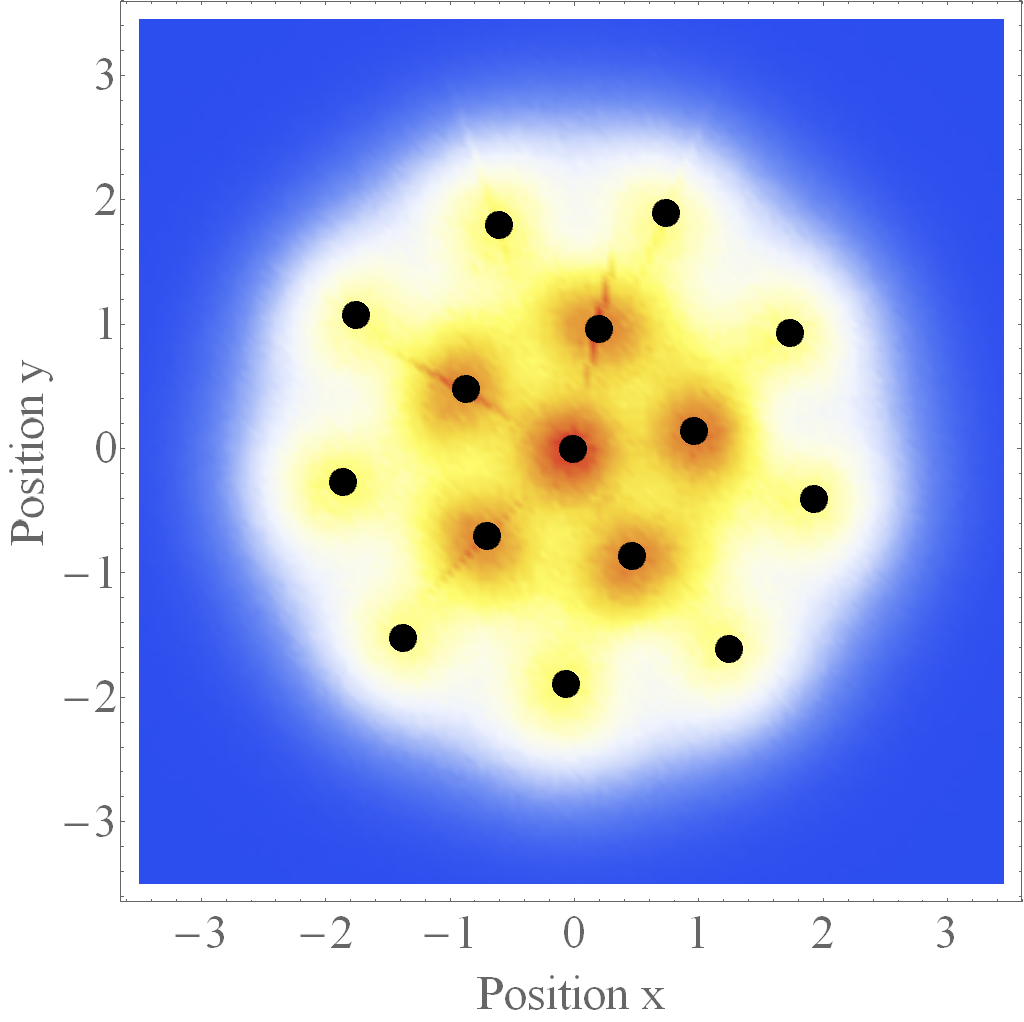}
   \includegraphics[height=6cm]{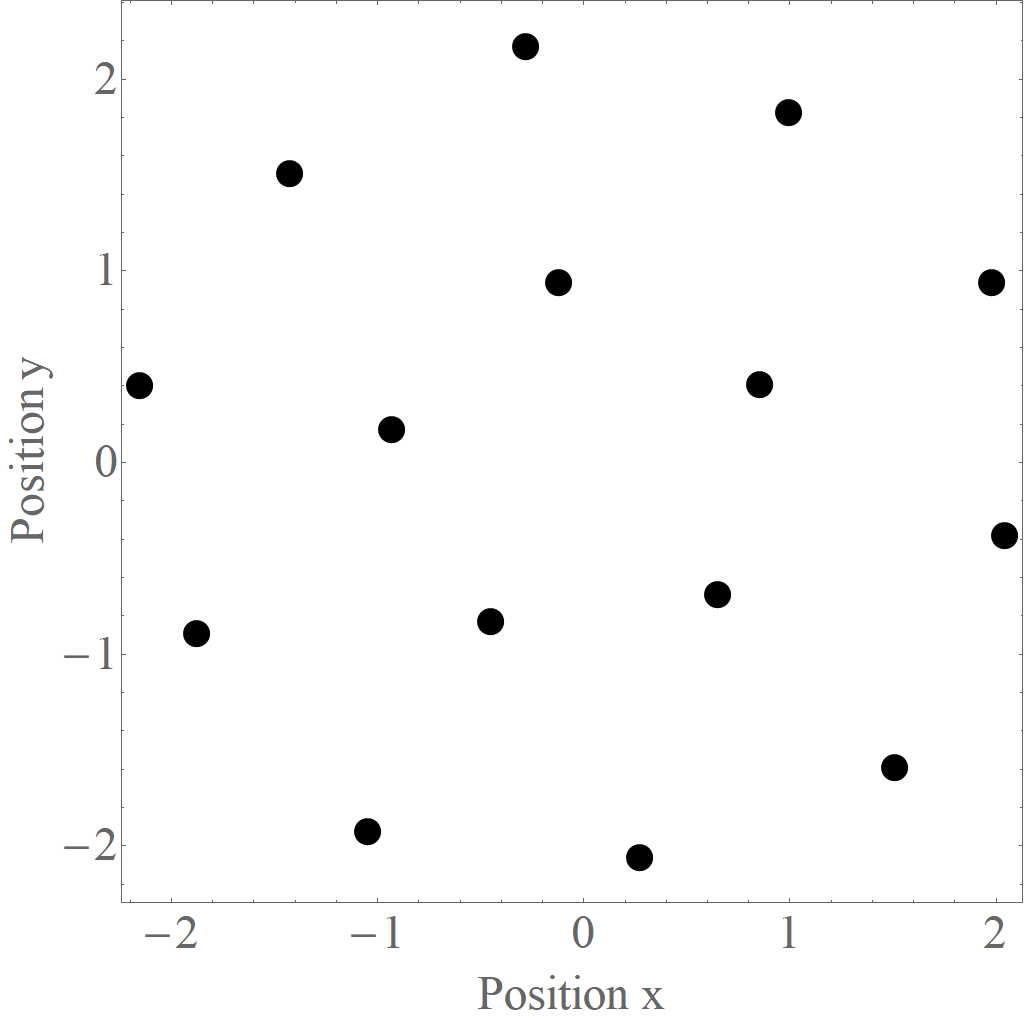}
  \caption{The Pauli crystal (left) and the Coulomb crystal (right) for $N=15$ particles confined in the isotropic harmonic trap. Colors and black dots correspond to the configuration density function ${\cal C}(\boldsymbol{r})$ and the most probable configuration of particles, respectively. All positions are scaled with respect to the natural oscillator length unit $\sqrt{\hbar/m\Omega}$.}
  \label{Fig2}
\end{figure}

\begin{figure}[H]
  \centering 
  \includegraphics[width=3cm]{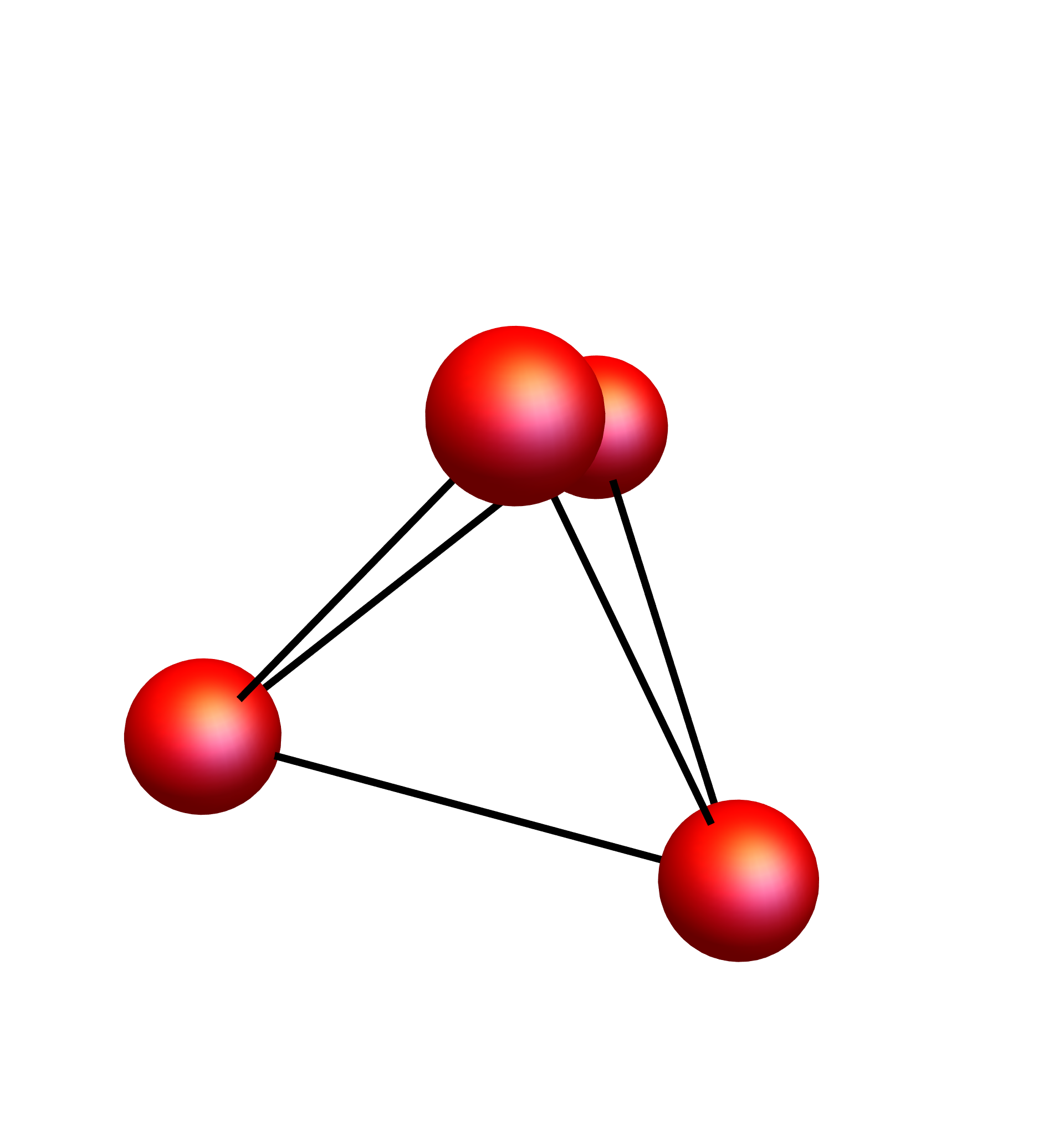}
  \includegraphics[height=4cm]{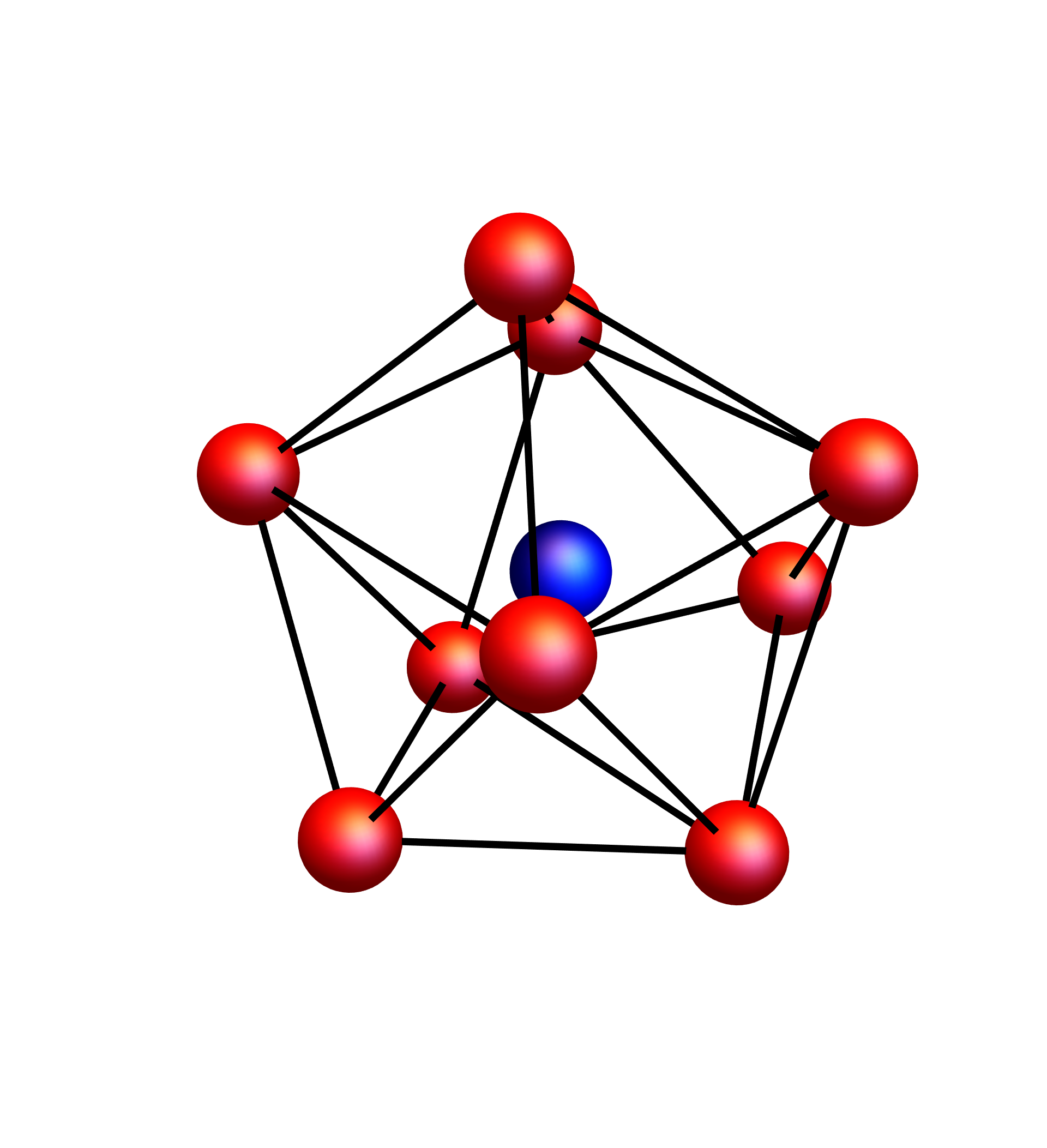}
  \includegraphics[height=5cm]{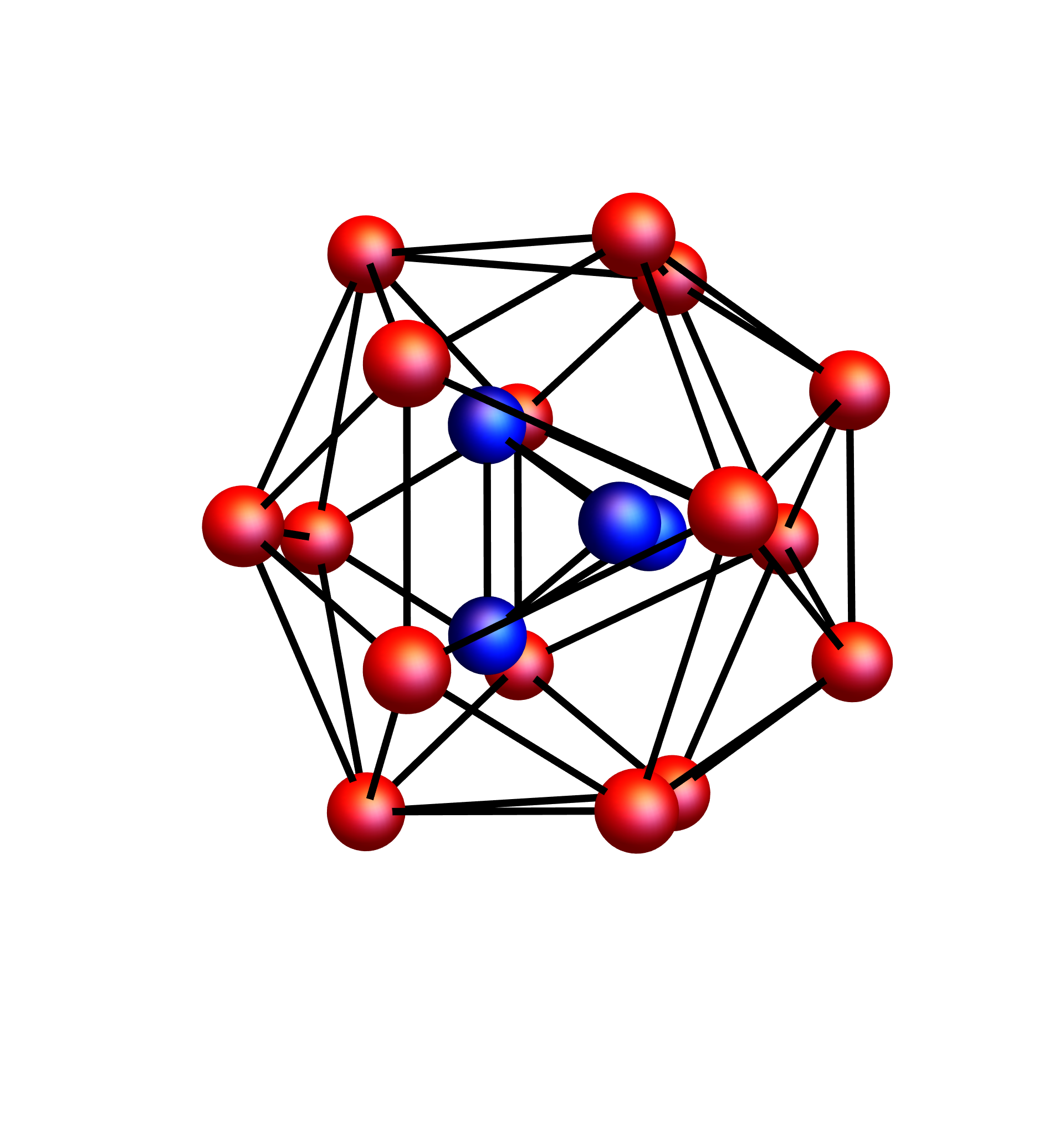}\vspace{-39pt}\\
  \includegraphics[width=3cm]{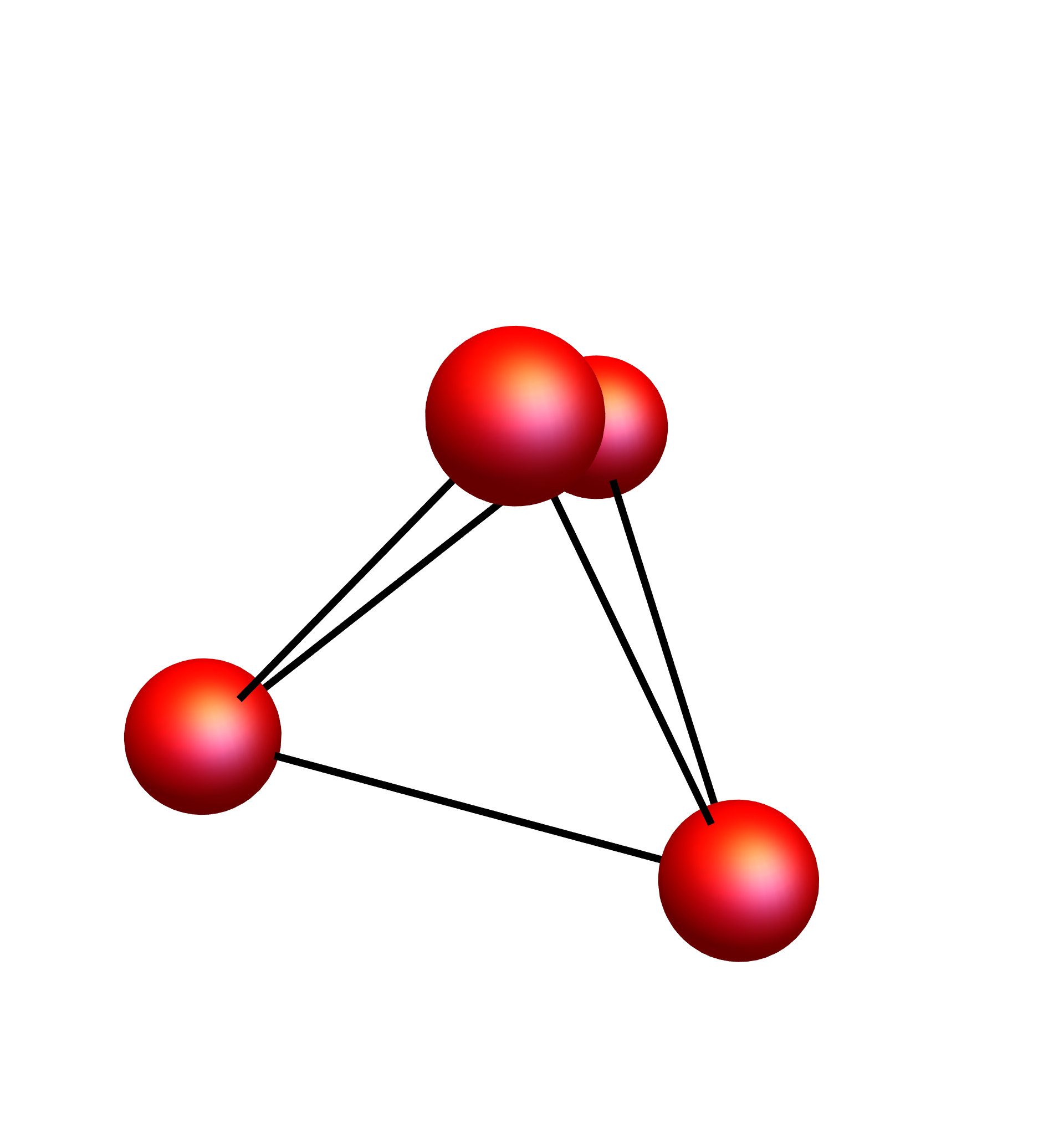}
  \includegraphics[height=4cm]{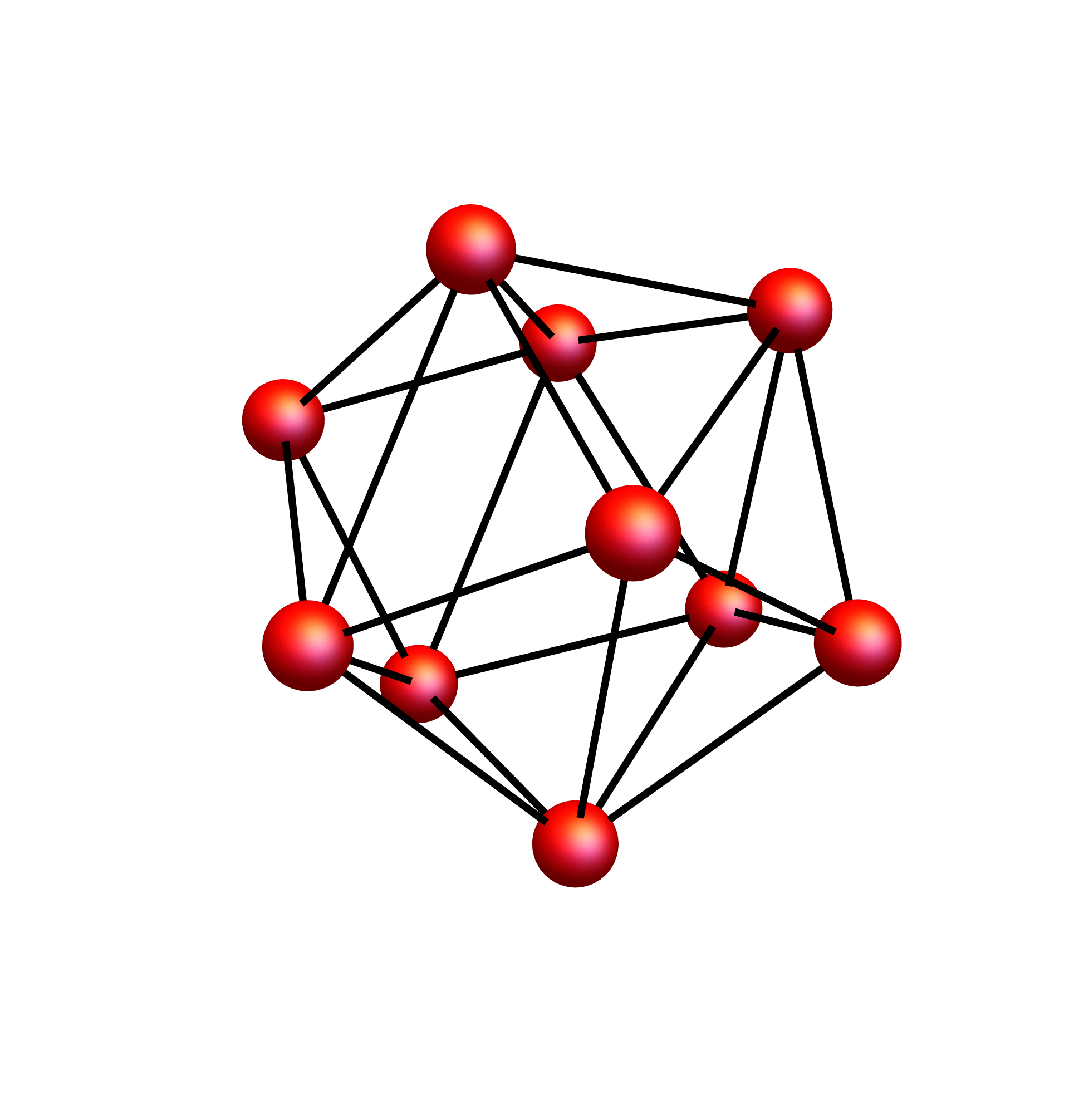}
  \includegraphics[height=5cm]{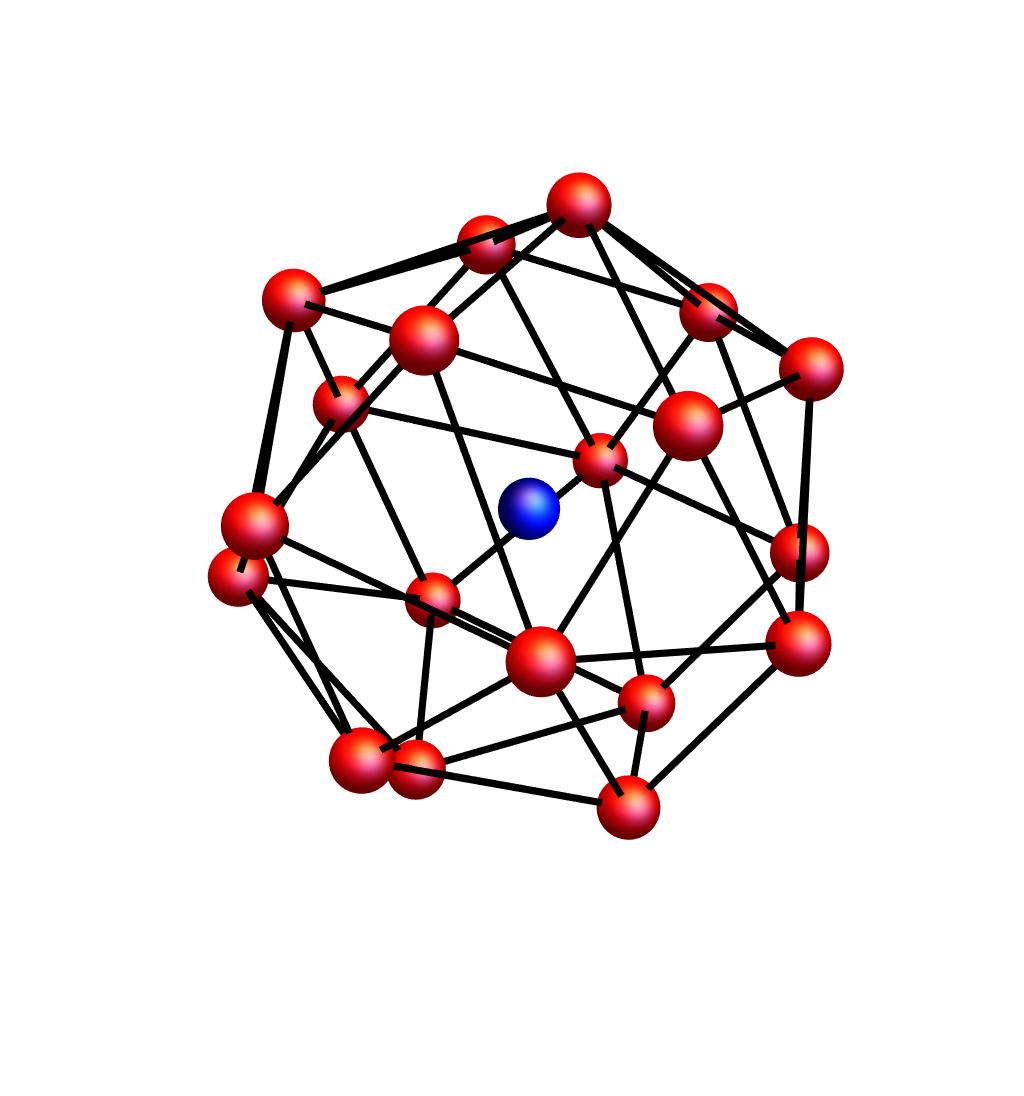}
  \caption{Comparison of three dimensional Pauli crystals (top panel) and Coulomb crystals (lower~panel) for $N=4$, $10$ and $20$ particles. Notice essential differences between the structures for 10 and 20~particles.}
  \label{Fig3}
\end{figure}

Inspection of $N=6$ particles case may suggest that the Pauli exclusion principle can be considered to be a specific two-body repulsive interaction. This however is not true and the result obtained for $N=6$ is not universal. To see this let us consider the case of $N=15$ particles (Figure~\ref{Fig2}). Here the configuration of the Pauli crystal consists of three geometric shells, containing respectively 1, 5, and 9 particles. The geometry of the inner shell exhibits fivefold symmetry imposed by the closest packing, the outer shell does not have this symmetry. The structure of the Coulomb crystal with the same number of particles, also shown in Figure~\ref{Fig2} for comparison, is essentially different. The Coulomb crystal contains only two geometrical shells, with $5$ and $10$ particles.  
The difference between the Pauli and Coulomb crystals can also be seen in three dimensions. To~clarify this, in Figure~\ref{Fig3} we compare shapes of Coulomb crystals with the corresponding Pauli crystals for $N=4$, $10$, and $20$ particles (for clearness we present the most probable configuration only). Both crystals have the same geometry in the case of $N=4$ particles. For $N=10$ particles, however, the Coulomb crystal contains two geometric shells--- the inner shell with one particle and the outer shell with 9 particles. For $N=20$ particles, both structures contain two geometric shells, but the inner shell of the Pauli crystal contains 4 particles, while of the Coulomb crystal only 1 particle. These examples clearly show that Pauli crystals are not governed by the closest packing principle. Thus, the exclusion principle cannot be considered to be a kind of two-body repulsive interactions. 

There is one more difference between the Pauli crystals and structures formed by atoms interacting by two-body forces. The probability distribution function for atomic positions, given by Equation (\ref{distribution}), exhibits one global maximum (defined up to symmetry transformations) regardless of the number of particles. This should be opposed to the energy function, which leads to equilibrium positions of charged particles. The latter has many local minima even for a relatively small number of particles~\cite{2005Coulomb}. This is one more evidence that the Pauli exclusion principle should not be considered to be a kind of mutual interaction of particles.  

An open question remains as to whether the exclusion principle can be modelled by repulsive many body interactions. It is clear that it cannot be true as a general statement. However, in some cases such modelling can lead to correct results, especially when more sophisticated approaches such as statistical interaction potential are considered~\cite{2017BatleAnnPhys,2019CiftjaAnnPhys}. 

\section{Open Energy Shells} 
In this section, we will extend our analysis to the open energy shells scenario. In these cases, the~ground-state energy does not determine the state uniquely since the most excited particles may occupy different orbitals. Obviously, the one-particle density $\rho^{(1)}(\boldsymbol{r})$ as well as the configuration density function ${\cal C}(\boldsymbol{r})$ (the shape of the Pauli crystal) depend on the whole state and not only on the energy. Thus, there is no unique structure of Pauli crystals for a given energy of the system.  
We will now discuss some examples.

\begin{figure}[H]
  \centering
  \includegraphics[height=5cm]{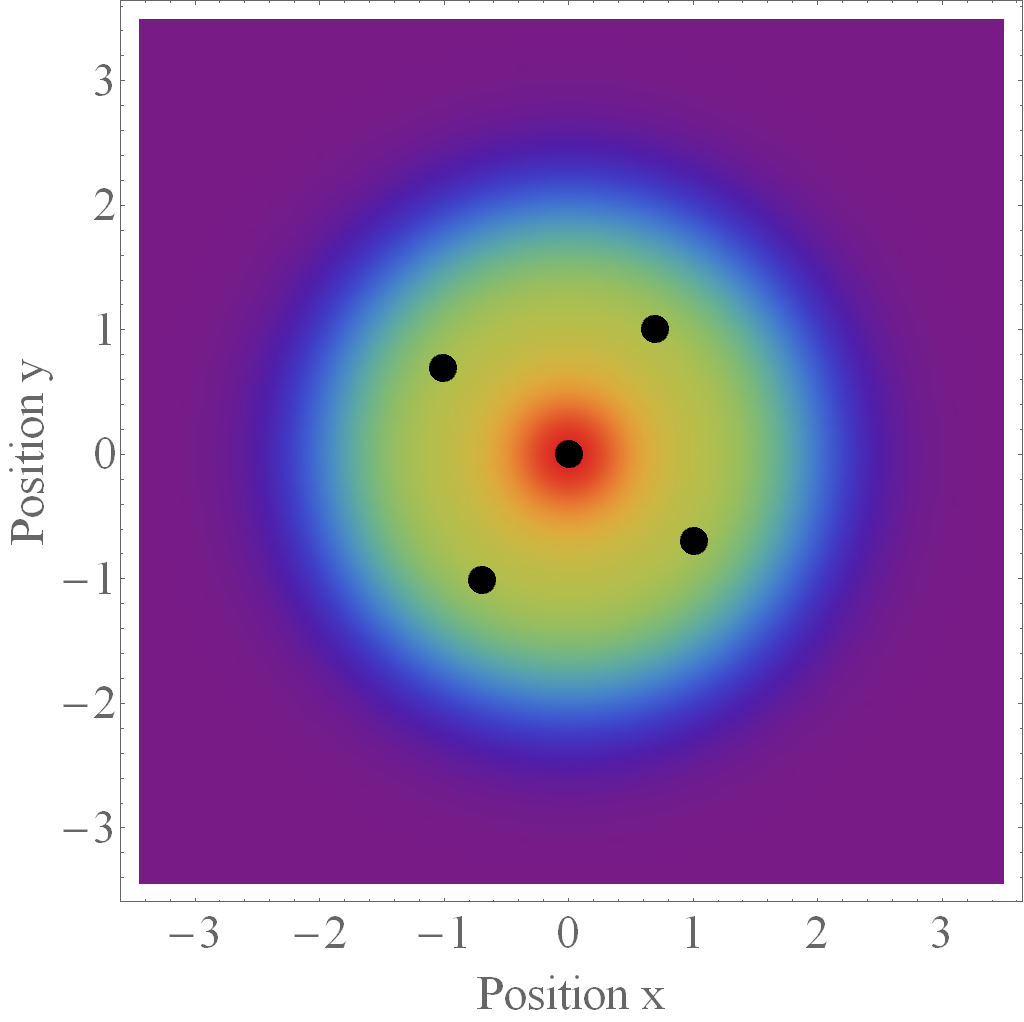} 
  \includegraphics[height=5cm]{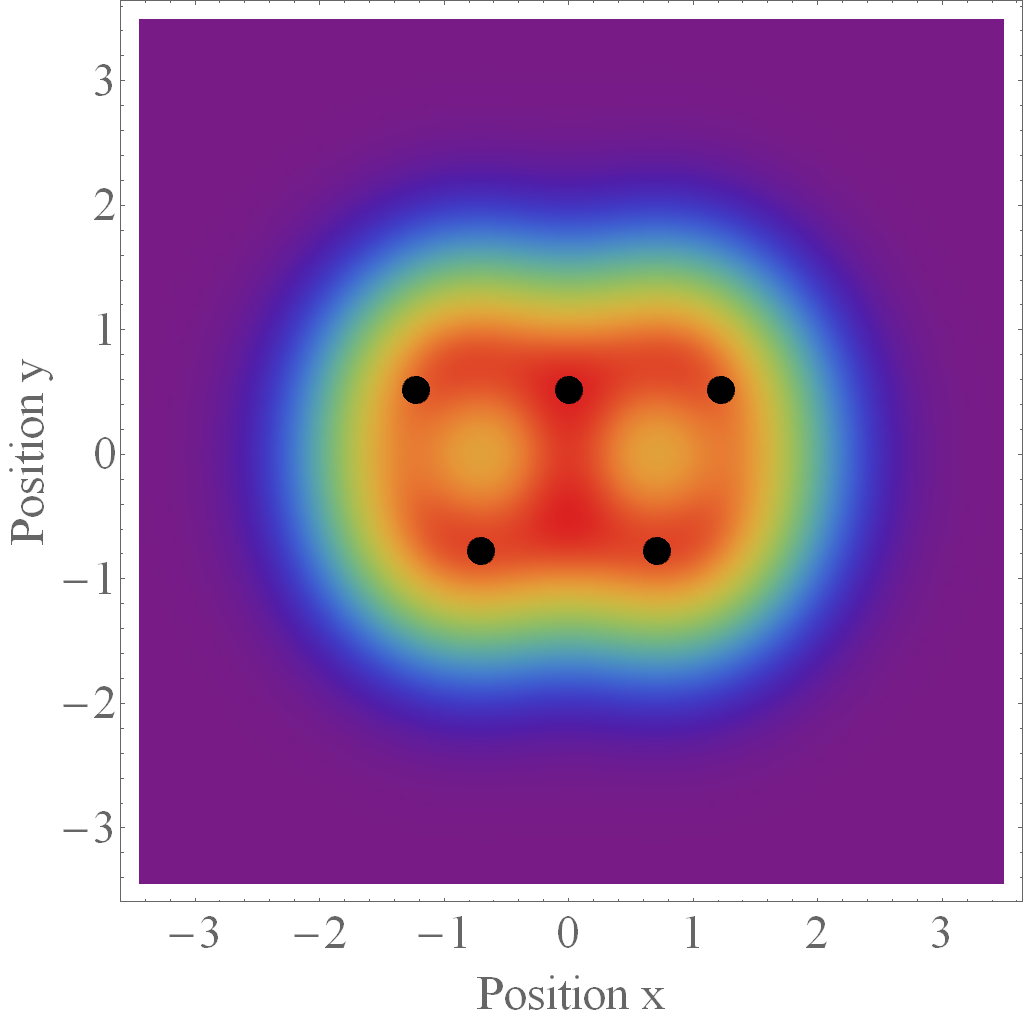}\\
  \includegraphics[height=5cm]{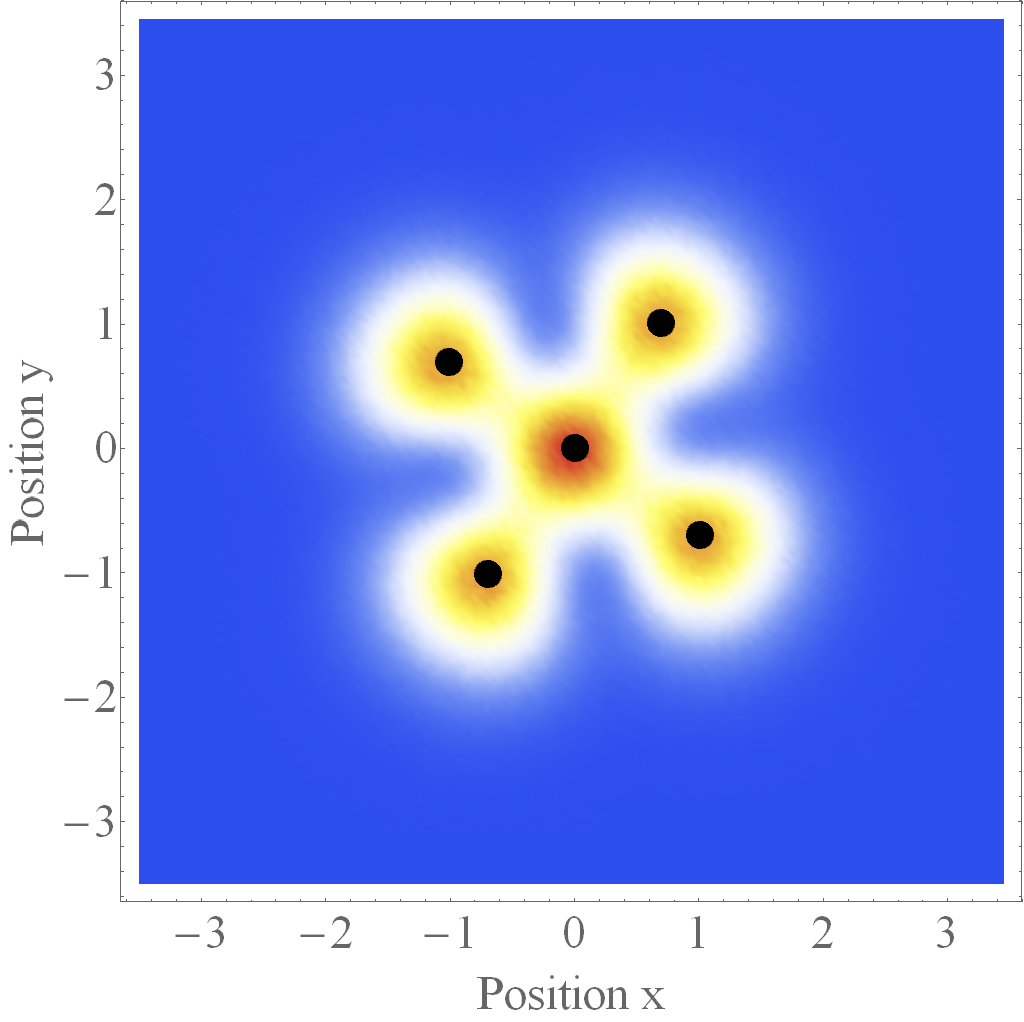}
  \includegraphics[height=5cm]{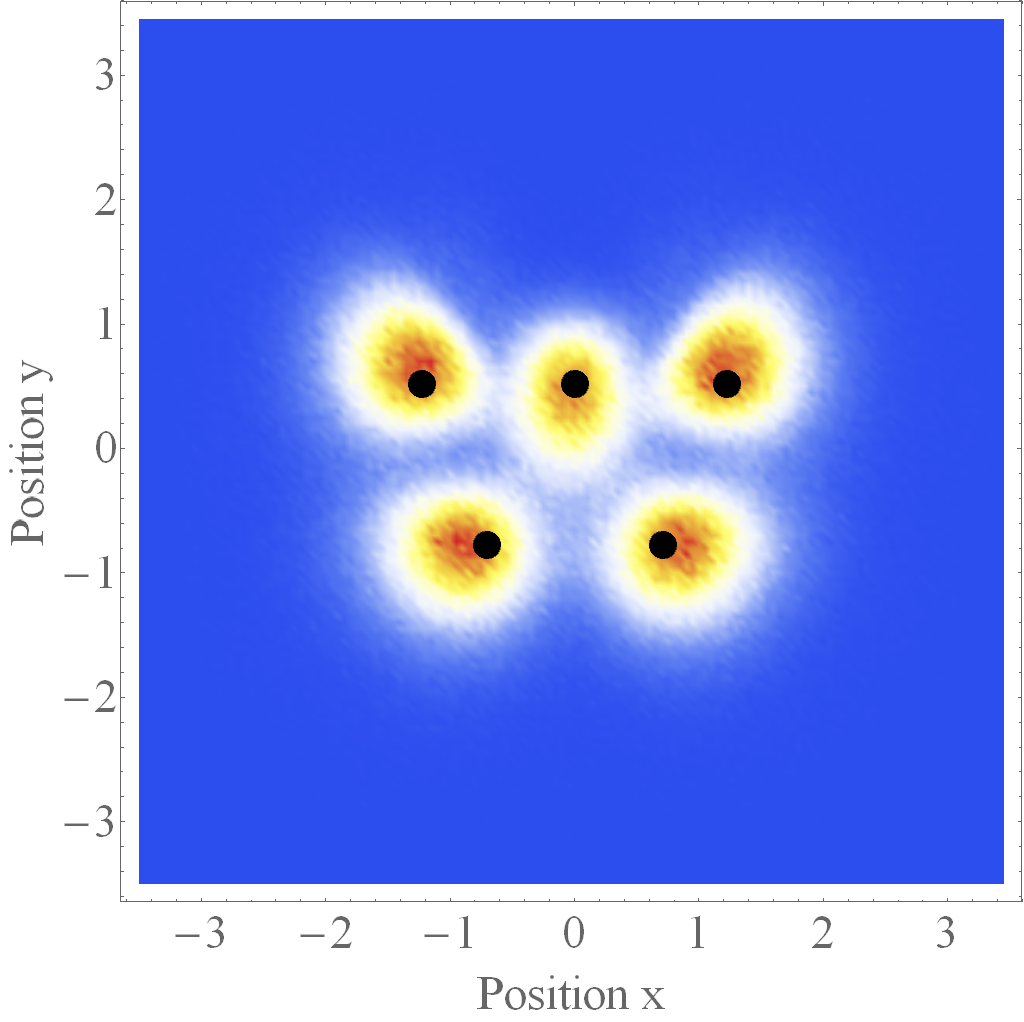}
  \caption{The one-particle density function ${\cal C}(\boldsymbol{r})$ (top) and the corresponding configuration density function ${\cal C}(\boldsymbol{r})$ (the Pauli crystal) (bottom) for $N=5$ particles confined in an isotropic harmonic trap. Left and right panels correspond to the circular (\ref{circstate}) and cartesian (\ref{cartstate}) states, respectively. See the main text for details. Black dots indicate the most probable configuration (the pattern). All positions are scaled with respect to the natural oscillator length unit $\sqrt{\hbar/m\Omega}$.}  \label{Fig4}
\end{figure}

First, let us consider the case of $N=5$ atoms (Figure~\ref{Fig4}). The first two energy shells (containing one and two particles, respectively) are fully occupied but the remaining two atoms may be distributed among three possible states. Among all possibilities, we will consider two specific examples. The~first uses the one-particle states with a given angular momentum to build the many particle state. We~assume that the levels with magnetic quantum numbers $m=-2$ and $m=0$ in the highest energy shell are occupied. The state, called the ''circular state'', is thus
\begin{equation} \label{circstate}
|\mathtt{S}\rangle= \hat{b}_{00}^\dagger \hat{b}_{01}^\dagger \hat{b}_{0-1}^\dagger \hat{b}_{0-2}^\dagger \hat{b}_{2 0}^\dagger |\mathtt{vac}\rangle 
\end{equation}
where lower indices denote the radial and angular momentum excitation respectively.
This state can be reached by introducing a small nonzero magnetic field in the trap, the field does not affect energies of the inner energy shells but removes the degeneracy of the one-particle states in the outer energy shell. In the ground state of the system, the one-particle states with the lowest energies are occupied. Finally, the magnetic field is set to zero, energies become degenerate but the wavefunction remains unchanged. In this way, we have selected a possible many-particle state. 
The shape of the Pauli crystal in such a case is shown in the top panel in Figure~\ref{Fig4}.
In the second example, we use the cartesian basis, namely the states in the outer energy shell are chosen as double excitation along the $x$ axis, single excitation along both $x$ and $y$ axes, and double excitation along the $y$ axis. We assume that the excitations in the $y$ direction have slightly higher energy than in the $x$ direction. Now, occupied one particle states have double excitation along the $x$ axis and single excitation along both axes. Finally, as before, the states are made degenerate, but the wave function remains unchanged. The many-particle state, called the “cartesian state” is
\begin{equation} \label{cartstate}
|\mathtt{S}\rangle= \hat{a}_{0 0}^\dagger \hat{a}_{0 1}^\dagger \hat{a}_{1 0}^\dagger \hat{a}_{2 0}^\dagger \hat{a}_{1 1}^\dagger |\mathtt{vac}\rangle 
\end{equation}

The Pauli crystal corresponding to this state is shown in the bottom panel in Figure~\ref{Fig4}, which differs from the previous shape. Other states describing an open shell exist and can be easily formulated.  

An important feature of the last case is that the maxima of the one-particle density functions do not coincide with the positions of particles forming the Pauli crystal. This is an illustration of a more general feature--- the one-particle density function does not give reliable information about positions of individual particles measured in a single measurement. One particle density gives averages of positions rather than information about the single measurements.

Let us note that the Pauli crystals do not have the symmetry of the trap, in the case of closed energy shells the symmetry is broken by the process of measurement. The Pauli crystals corresponding to open energy shells break the trap symmetry in a much stronger way, measurements are additional factors leading to an even stronger violation. The two examples described above show clearly that the symmetry of the Pauli crystal is lower than the symmetry of the trap. In the first case considered above the crystal has the rotational symmetry, while in the second case the Pauli crystal is not rotationally~symmetric.

At this point, it should be emphasized that the mentioned degeneracy of open-shell ground states cannot be captured by any semi-classical method which treats quantum particles obeying the Pauli exclusion principle as a kind of classical system with appropriately tailored mutual interactions. All features caused by the degeneracy of the many-body ground state are solely reserved for a fully quantum description of the system.

\section{Particles in a Square Potential Well}

\begin{figure}[H]
  \centering
  \includegraphics[height=4.5cm]{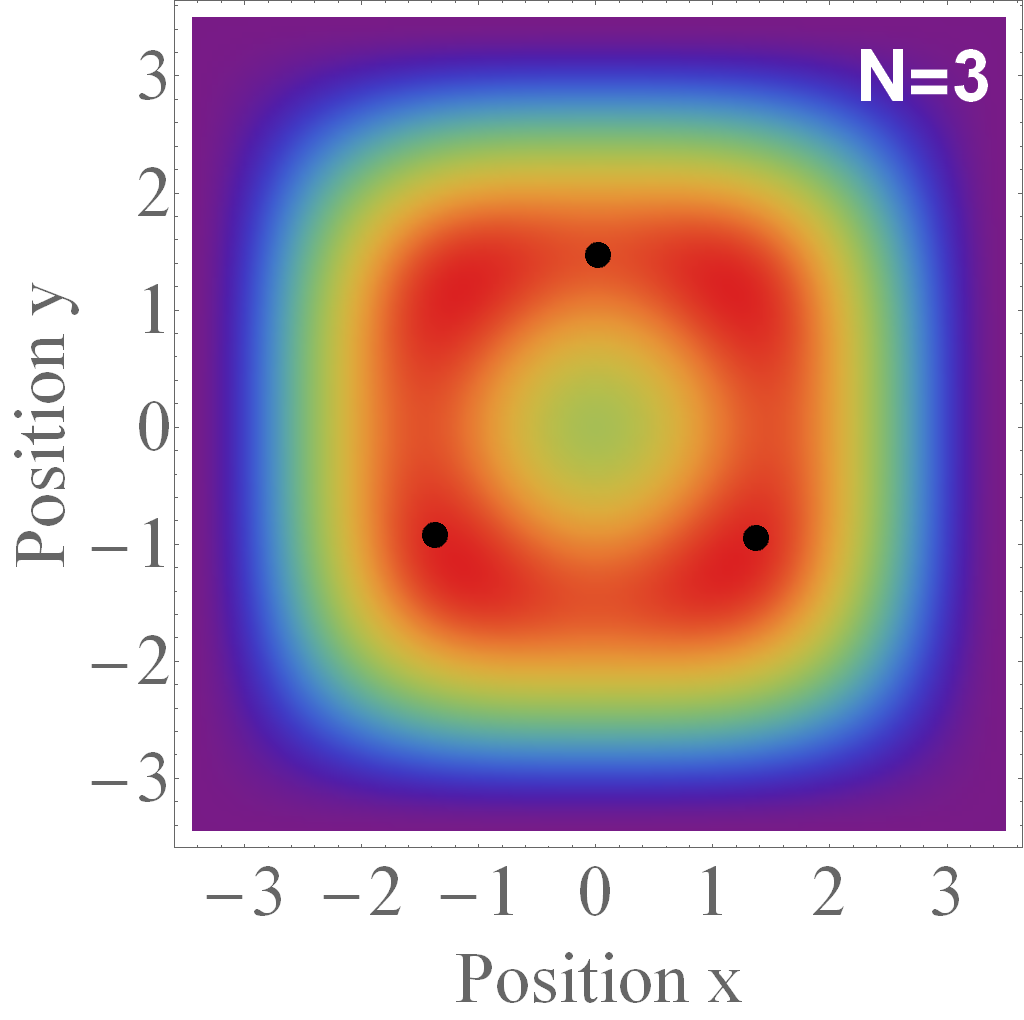}
  \includegraphics[height=4.5cm]{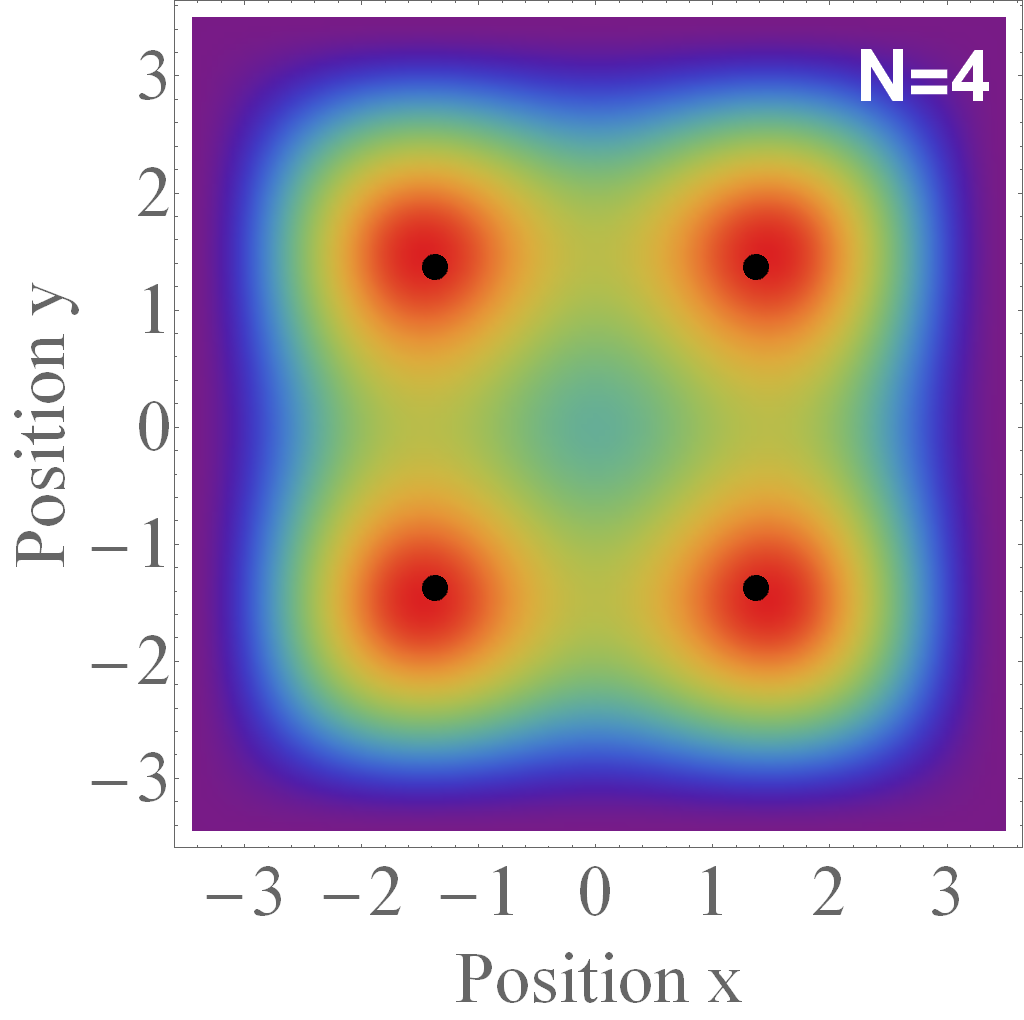}
  \includegraphics[height=4.5cm]{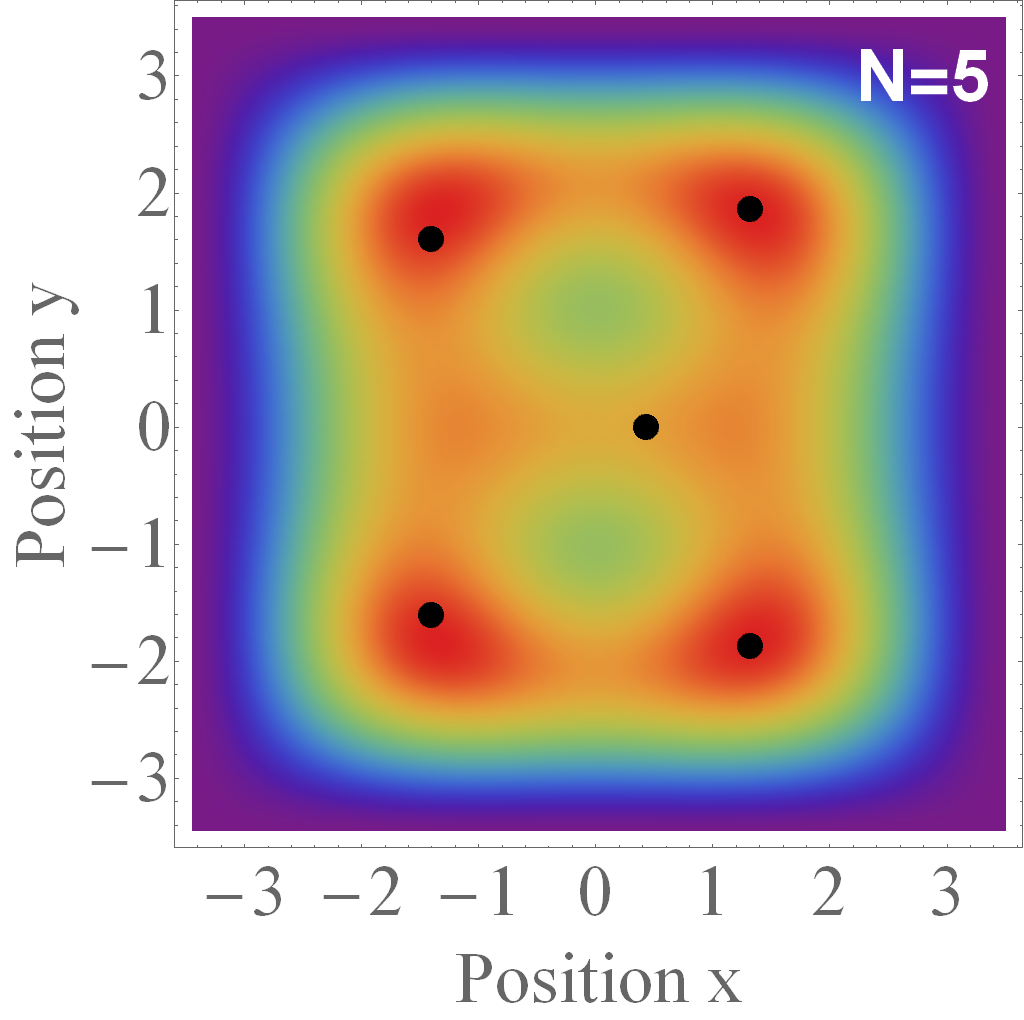}\\
  \includegraphics[height=4.5cm]{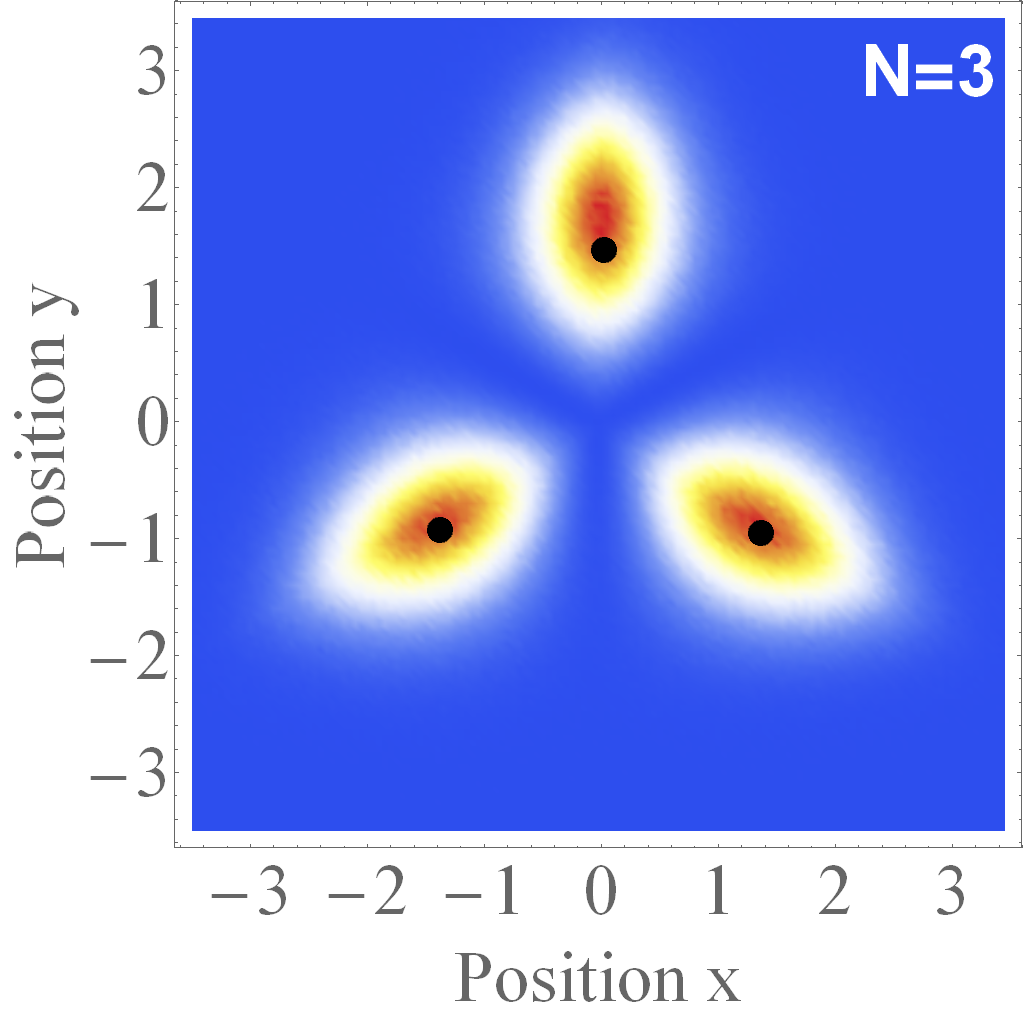}
  \includegraphics[height=4.5cm]{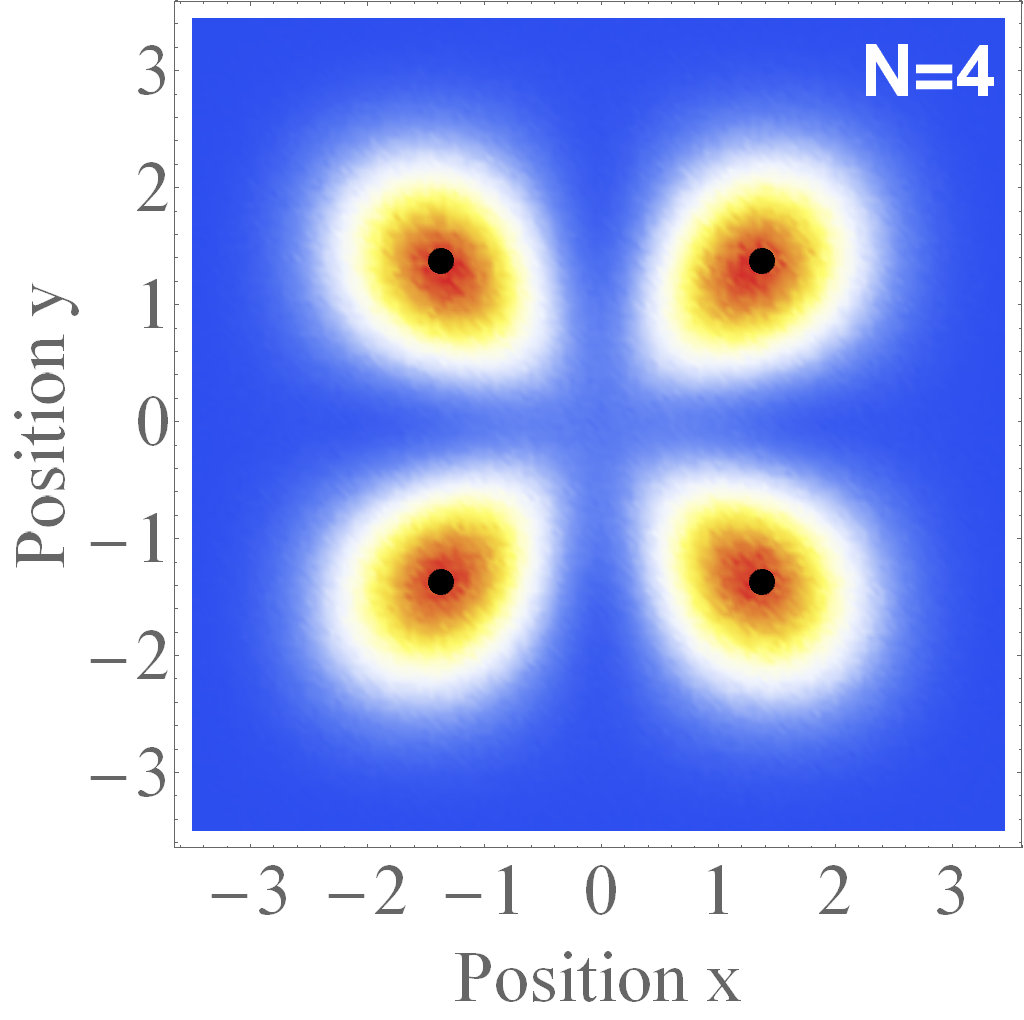}
  \includegraphics[height=4.5cm]{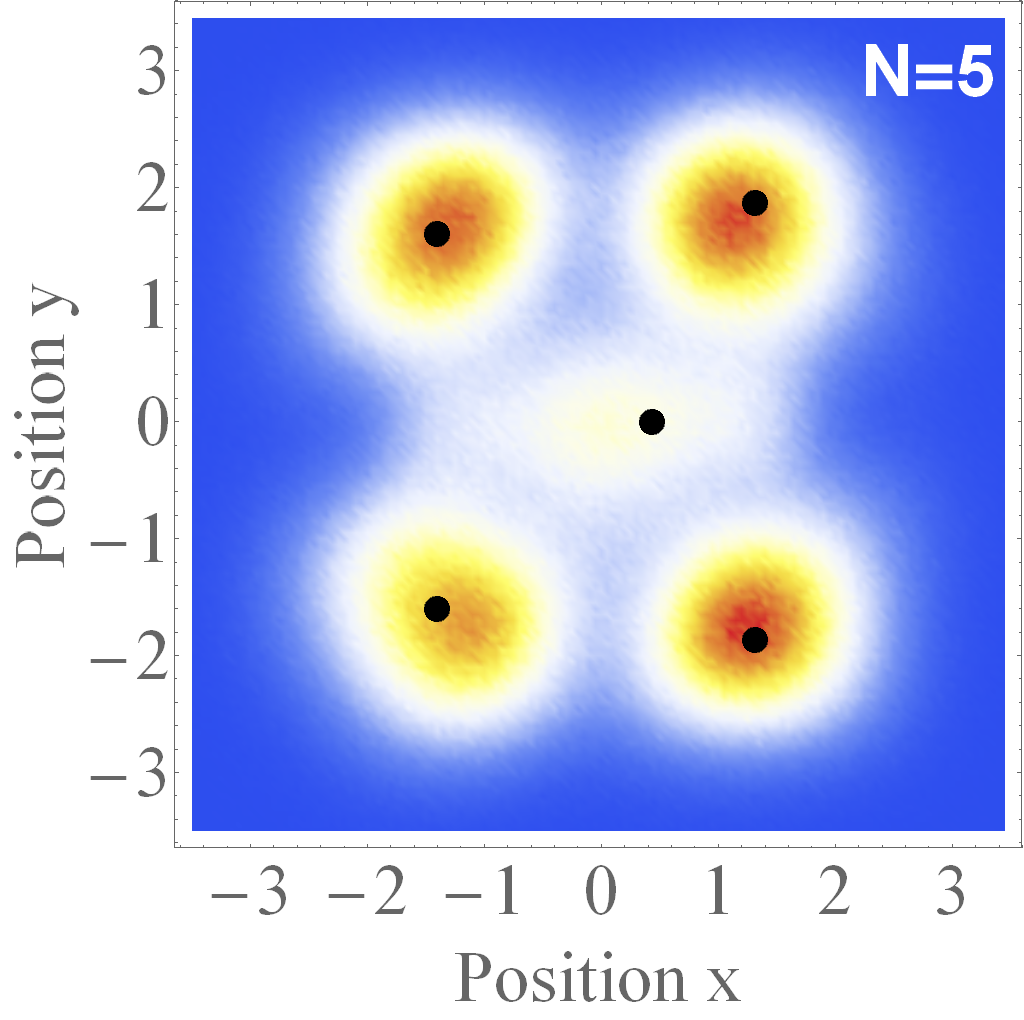}\\
  \includegraphics[height=4.5cm]{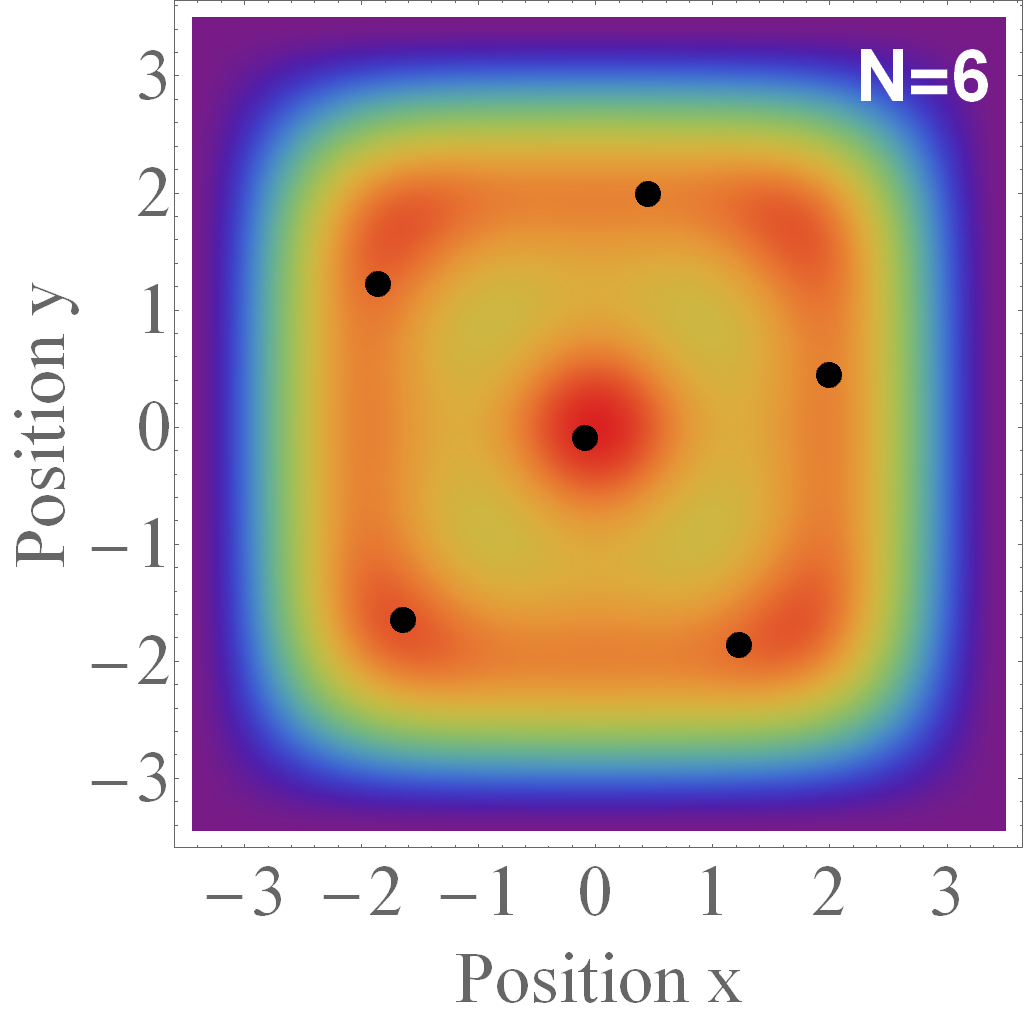}
  \includegraphics[height=4.5cm]{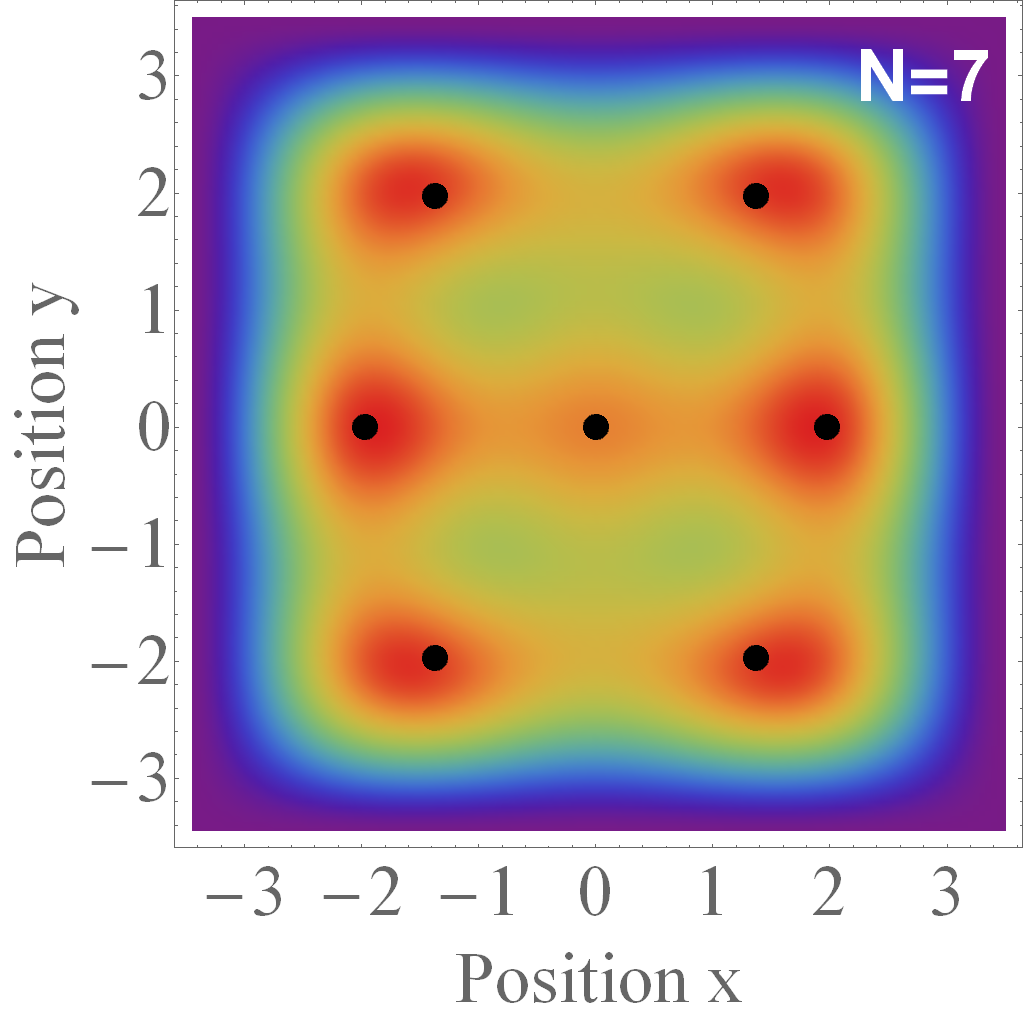}
  \includegraphics[height=4.5cm]{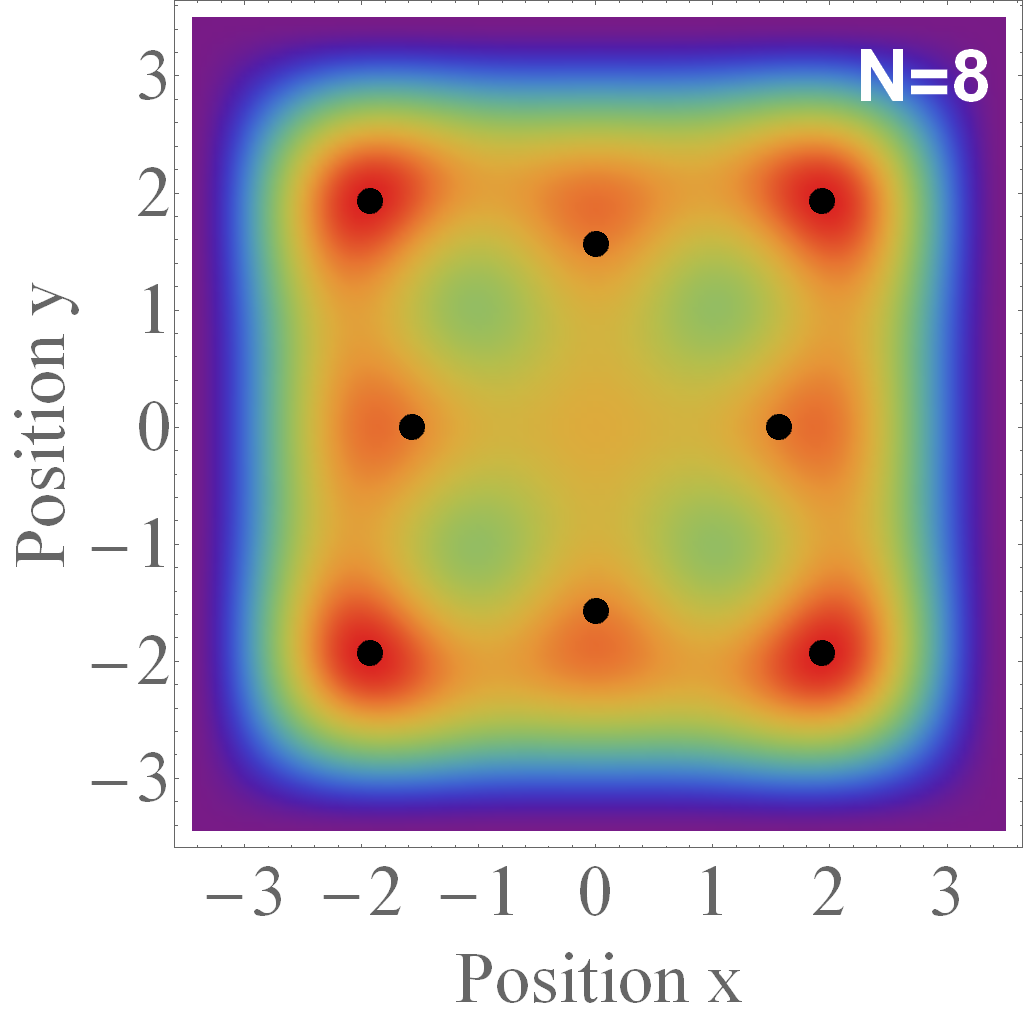}\\
  \includegraphics[height=4.5cm]{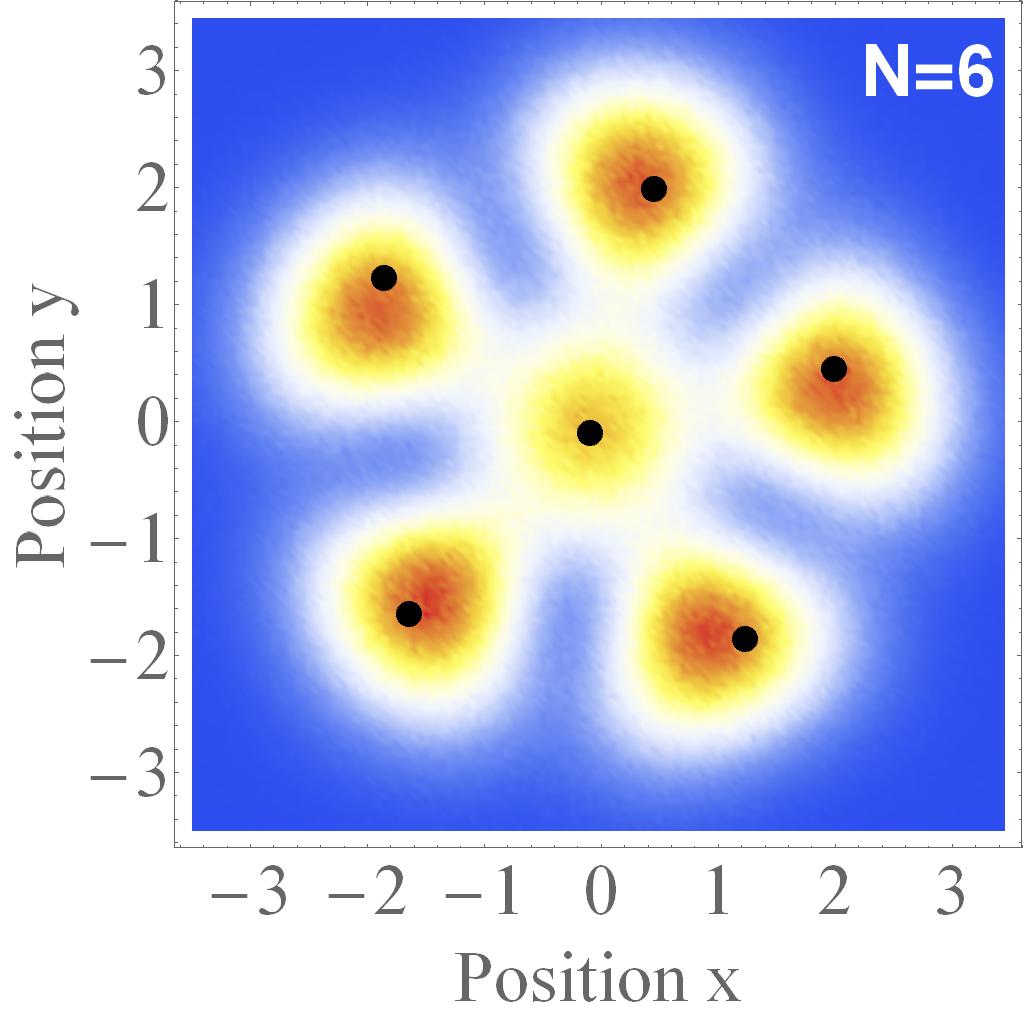}
  \includegraphics[height=4.5cm]{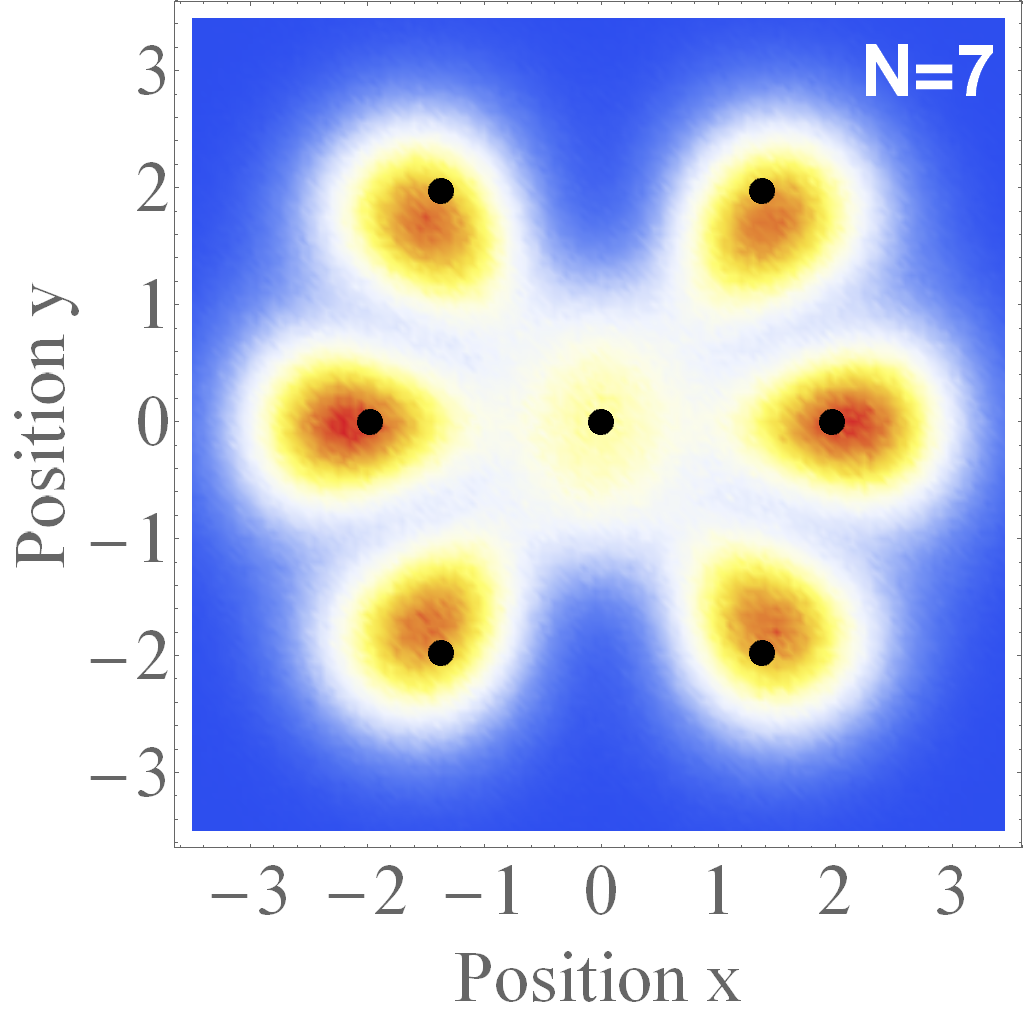}
  \includegraphics[height=4.5cm]{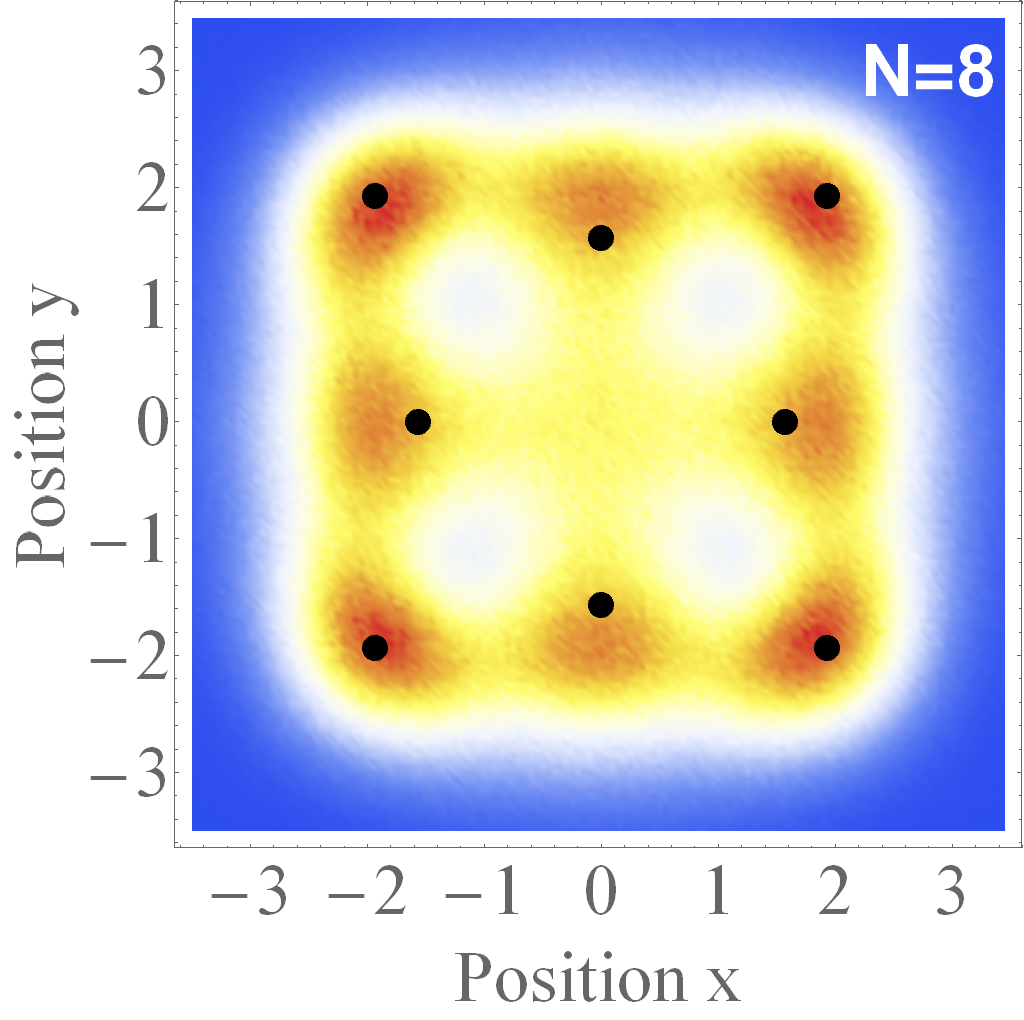}
  \caption{One-particle density function (first and third row) and corresponding configuration density function ${\cal C}(\boldsymbol{r})$ exposing Pauli crystals (second and fourth row) for $N=3,\ldots,8$ particles in a square well trap. Black points denote the most probable configuration of the many-body distribution $\rho^{(N)}$. Notice~that the maxima of one particle density not always coincide with the position of particles forming the Pauli crystal. All positions are scaled with respect to the size of the well $a$.}
  \label{Fig5}
\end{figure}

In this section, we will discuss the case when the particles are bound by a two-dimensional square well potential. Thus, the external potential has a lower symmetry, the natural fivefold symmetry of the particles competes with the lower symmetry of the potential. Denoting the size of the well by $a$, normalized one-particle wave functions have a form:
\begin{equation}
\psi_{nm}(x,y)=\frac{2}{a}\sin\left(n\frac{\pi x}{a}\right)\sin\left(m\frac{\pi y}{a}\right),
\end{equation}
where $n$ and $m$ are natural numbers denoting spatial excitations in $x$ and $y$ direction, respectively. The~operators creating a particle in the state with wave function $\psi_{nm}(x,y)$ are denoted by $\hat{c}_{nm}^\dagger$. As~before, first we find maxima of the many-particle probability distribution function $\rho^{(N)}(\boldsymbol{r}_1,\ldots,\boldsymbol{r}_N)$ in the many-body ground-state. Then we simulate single-shot measurements of positions of all particles and generate the configuration density function ${\cal C}(\boldsymbol{r})$. Since the symmetry of the potential is now lower than in the previous examples, therefore the distillation has to be restricted to transformations that do not violate the symmetry of the potential. The results are shown in Figure~\ref{Fig5}. The top row shows the one-particle density function $\rho^{(1)}(\boldsymbol{r})$ for $N=3,\ldots,8$ particles, while in the lower row, the~corresponding configuration density function ${\cal C}(\boldsymbol{r})$ is displayed. The simple case of $N=3$ particles is very interesting. Let us note that in this case, the particles occupy the first two energy shells and that these shells are closed. While the particles themselves prefer the triangular symmetry, the external potential has the symmetry of a square. As a result, the one-particle density distribution $\rho^{(1)}(\boldsymbol{r})$ has the symmetry of the potential and exhibits no sign of three maxima. On the contrary, the configuration density function ${\cal C}(\boldsymbol{r})$ has purely triangular symmetry and the positions of three particles are clearly distinguished. In this case, the exclusion principle dominates over the symmetry of the potential and consequently leads to the triangular configuration of the Pauli crystal.

The situation is essentially different in the case of $N=4$ particles. The next energy shell is not fully occupied and the fourth particle is doubly excited. Assuming that this double excitation is along the $x$ axis the many-body ground state of the system is
\begin{equation}
  |\mathtt{S}\rangle= \hat{c}_{0 0}^\dagger \hat{c}_{0 1}^\dagger \hat{c}_{1 0}^\dagger \hat{c}_{2 0}^\dagger |\mathtt{vac}\rangle,
\end{equation}

As clearly seen in Figure~\ref{Fig5} (the second column), in this case the maxima of the one-particle density function $\rho^{(1)}(\boldsymbol{r})$ mirror the shape of the most probable configuration and the shape of the Pauli crystal. The whole system has the symmetry of the square--- the symmetry of the potential. This is, however, an exceptional case, as illustrated by the state with $N=5$ particles. For example, if one considers a five-particle state of the form
\begin{equation}
  |\mathtt{S}\rangle= \hat{c}_{0 0}^\dagger \hat{c}_{0 1}^\dagger \hat{c}_{1 0}^\dagger \hat{c}_{2 0}^\dagger \hat{c}_{1 1}^\dagger|\mathtt{vac}\rangle
\end{equation}
then the resulting single-particle density $\rho^{(1)}(\boldsymbol{r})$ and the configuration density function ${\cal C}(\boldsymbol{r})$ have rectangular rather than square shape (the third column in Figure~\ref{Fig5}). Closer inspection of the~results shows that the one-particle density function only vaguely resembles the configuration density. However, the configuration density does not have the symmetry of the potential. 

The third energy shell is finally closed in the case of $N=6$ particles and the Pauli crystal forms two geometric shells (the fourth column in Figure~\ref{Fig5}). The inner geometric shell contains one particle, the outer one contains the remaining five particles, such as in the case of a harmonic isotropic trap. Observe, however, that the one-particle density function $\rho^{(1)}(\boldsymbol{r})$ exhibits only four maxima in the outer geometric shell (roughly in the corners of the trap). In contrast, the configuration density ${\cal C}(\boldsymbol{r})$ shows clearly that the outer geometric shell contains five particles. Similar features are seen in the case of $N=7$ particles. These effects illustrate a general statement that the shape of the one-particle density profile $\rho^{(1)}(\boldsymbol{r})$ does not always give proper positions of the particles if they are measured simultaneously. An exactly opposite mismatch between the number of particles and the number of maxima in the one-particle density is seen in the case of $N=8$ particles. Namely, the one-particle density function $\rho^{(1)}(\boldsymbol{r})$ has nine maxima on the outer geometric shell, while the Pauli crystal contains only eight particles there. A clear local maximum of the one-particle density function located in the center of the potential does not correspond to the presence of a particle.

These examples show clearly the interplay between the symmetry of the external potential with the symmetry imposed by the quantum statistics. While the fermionic nature of particles supports rather fivefold symmetry, the potential supports the symmetry of the square. While the shape of the one-particle density function $\rho^{(1)}(\boldsymbol{r})$ is rather dominated by the symmetry of the potential, the shape of the configuration density function ${\cal C}(\boldsymbol{r})$ is quite difficult to predict and emerges as a result of non-trivial competition between the quantum statistics and the external confinement.

\section{Conclusions}
We discussed possible configurations of particles with the fermi statistics in a confining potential. The Pauli exclusion principle prevents particles from being close to each other even if the particles do not interact. Thus, the exclusion principle leads to correlations between particle positions. We~pointed out that the particles tend to take positions in vertexes of non-trivial polygons, called the Pauli crystals. We compared positions of these vertexes with maxima of the one-particle density functions and showed that in many cases they do not coincide. Thus the maxima of the one density distribution function cannot be identified with the most probable positions of particles.

Positions of particles form geometrical structures, we compared symmetries of these structures with the symmetries of the confining potential. It turned out that there is no simple relation between these two symmetries. Rather the shape of the Pauli crystals results from an interplay between the symmetry of the potential and the natural symmetry of fermions. 
 
We compared the Pauli crystals with analogous structures formed by interacting particles, namely electrically charged particles in a trap (Coulomb crystals). We showed that although the shapes are the same for a small number of particles they become different for a larger (larger than 8) particle number. This is a clear indication that the Pauli principle cannot be considered to be a kind of interaction. Rather it provides a mechanism leading to high order correlations between the particles that in general cannot be modeled by two-body interactions.
\vspace{6pt}

\authorcontributions{All the authors contributed equally. All authors have read and agreed to the published version of the manuscript.}

\funding{This research was funded by the (Polish) National Science Center Grants No. 2017/25/B/ST2/01943 (MP), 2019/32/Z/ST2/00016 (MG), and 2016/22/E/ST2/00555 (TS).}

\acknowledgments{The authors are grateful to Selim Jochim for very fruitful discussions.}

\conflictsofinterest{The authors declare no conflict of interest.} 

\reftitle{References}

\end{document}